## UX Heuristics and Checklist for Deep Learning powered Mobile Applications with Image Classification

Christiane Gresse von Wangenheim, Gustavo Dirschnabel
Department of Informatics and Statistics, Federal University of Santa Catarina, Florianópolis/SC, Brazil

**Abstract**
Advances in mobile applications providing image classification enabled by Deep Learning require innovative User Experience solutions in order to assure their adequate use by users. To aid the design process, usability heuristics are typically customized for a specific kind of application. Therefore, based on a literature review and analyzing existing mobile applications with image classification, we propose an initial set of AIX heuristics for Deep Learning powered mobile applications with image classification decomposed into a checklist. In order to facilitate the usage of the checklist we also developed an online course presenting the concepts and heuristics as well as a web-based tool in order to support an evaluation using these heuristics. These results of this research can be used to guide the design of the interfaces of such applications as well as support the conduction of heuristic evaluations supporting practitioners to develop image classification apps that people can understand, trust, and can engage with effectively.

## 1. Introduction

Today's AI advances are mainly based on Deep Learning (DL) which is a subfield of machine learning that involves training artificial neural networks to recognize patterns in data. These networks are composed of layers of interconnected nodes that can learn to identify features and patterns in the input data. During training, the weights of the connections between the nodes are adjusted to minimize the error in the network's predictions maximizing its accuracy. Once the network is trained, it can be used to make predictions on new, unseen data.

Deep learning has achieved state-of-the-art performance in many applications, including image classification. Image classification is a process in which a computer algorithm analyzes and categorizes images based on their visual features. It involves training a deep neural network on a large dataset of labeled images, where each image is associated with a specific class, e.g. a cat breed. The neural network learns to recognize patterns and features in the images that are indicative of their respective classes. Once trained, the algorithm can classify new images by comparing their features to the learned patterns, and outputting the predicted class label. The accuracy of the classification model depends on the quality and quantity of the training data, as well as the architecture and hyperparameters of the neural network.

The recent evolution of Deep Learning has also increased the deployment of these trained image classification models into mobile applications, in diverse domains, including computer vision. This allows the app to automatically make predictions based on the contents of the image, e.g., classifying dog breeds, plants, or even diseases such as skin cancer.

However, different to traditional software systems, there are various risks associated with interaction with DL-powered systems.

One of the main risks associated with such systems is accuracy. Due to the probabilistic nature of DL algorithms their results are not always definitive or 100% accurate. While DL-powered systems can perform tasks at a high level of accuracy, they also make errors, particularly when dealing with complex or ambiguous situations. This can lead to incorrect decisions or recommendations, which can have serious consequences, e.g., when classifying images of skin cancer.

DL-powered systems can also be opaque or difficult to understand, making it difficult for users to know how they are working or what data they are using. This lack of transparency can lead to distrust or suspicion among users, particularly when the system is making important decisions or providing recommendations. This situation is commonly referred to as the black box problem in Artificial Intelligence (AI). Without understanding how AI reaches its conclusions, it is an open question to what extent the user can trust these systems. The question of trust becomes more urgent as more and more decision-making is delegated to AI to areas that may harm humans, such as security, healthcare, and safety.

Results of DL-powered systems can also be unethical, such as in image classifications misclassifying people of a certain race due to algorithmic bias, where the DL model had been trained on a dataset that was not diverse enough to accurately classify images.

DL-powered systems can also pose risks to privacy. These systems may collect large amounts of data and/or images from the users, which can be used for purposes that the user may not be aware of or may not have consented to.

As a result, DL-infused systems may demonstrate unpredictable behaviors that can be disruptive, confusing, offensive, and even dangerous (Amershi et al., 2019). These risks can become even more critical when considering the larger audience of AI-illiterate general public using such mobile apps. Since AI/DL technology is still new and not well understood by many people, users may trust the app's decisions without questioning them. Inaccurate or biased results generated by the app can lead to false conclusions, misunderstandings, or even harm the user (e.g., when relying solely on a skin cancer application app, without consulting a dermatologist). The lack of understanding of how AI works can also make it difficult for users to know when to rely on the app's suggestions and when to question them. Therefore, it is crucial to consider the usability and transparency of AI-powered mobile apps to avoid potential harm and ensure user trust. This lack of transparency can create confusion, frustration and mistrust and indeed specific socially untoward consequences of algorithmic interactions have been widely identified (Carroll, 2022)(Ehsan et al., 2021)(Gaube et al., 2021)(Ribeira & Lapedriza, 2019).

Yet, as the deployment of DL in mobile applications grows, the field of user experience design must shift to understand the possibilities, limitations, and biases of AI. A great image classification app depends on a well-designed DL model as much as it depends on a well-designed user interface (UI) and user experience (UX) that compose the human experience around the AI models (van Allen, 2018). Aiming to enhance or complement human capabilities rather than to replace them through DL, a Human-Centered AI (HCAI) approach needs to be adopted to develop and deploy AI systems that focus on the needs and well-being of the individuals and communities affected by the technology (Rield, 2019)(Ribeira & Lapedriza, 2019)(Xu, 2019). HCAI aims to ensure that the AI technology is developed with the user's needs and expectations in mind, making it more accessible and easy to use. Human-centered design can help to mitigate the risks associated with AI. By prioritizing the user's experience and understanding of the technology, designers should develop interfaces that provide users with greater control and understanding of the AI's decisions and outputs (Shneiderman, 2020)(Xu, 2019).

As such, it is important that these systems provide an adequate user experience (UX) in order to be effective and easy to use. A well-designed UX can help ensure that people are able to use an AI system easily and effectively, and that they are able to get the most out of it and prevent any harm (Wong, 2018). Yet, the design of user interfaces of DL-powered apps still presents a challenge and limited attention has been given so far to the user interface design principles for such systems (Yang et al., 2020).

AI User Experience (AIX) It refers to the user experience design of AI-powered systems, with a focus on creating interfaces and interactions that are intuitive, transparent, ethical, usable, useful and trustworthy for users. The field of AIX is still relatively new, so far few proposals of guidance have emerged for AIX design (Wright et al., 2020), mostly in white papers by large technology companies such as Google’s People + AI Guidebook (Google, 2023) for any kind of AI systems with emphasis on recommendation systems or IBM’s AI Design Guidelines (IBM, 2023), while Amazon (2020) and Facebook (2023) propose guidelines for conversational AI. Amershi et al. (2019) at Microsoft synthesized research in interaction with AI into a set of guidelines for human-AI interaction (HAI guidelines) as well as Wright et al. (2020) presenting a comparative analysis of industry human-AI interaction guidelines. These frameworks either provide general principles and guidelines or focus more on a specific type of AI-powered system (such as recommendation systems or conversational agents) that are difficult to apply to other AI tasks, such as image classification.

Therefore, this article presents an initial proposal of AIX heuristics and checklist in order to provide guidance for designing and evaluating mobile apps with image classification powered by Deep Learning. The availability of such heuristics is expected to contribute to the improvement of the user interface design of such apps and, thus, providing an improved user experience.

## 2. Background

### 2.1 Image Classification Applications powered by Deep Learning

A prominent AI task today is image classification applied to various domains such as healthcare, biology, arts, etc. The image classification task predicts the class of an object in an image, e.g., a cat breed (Figure 1).

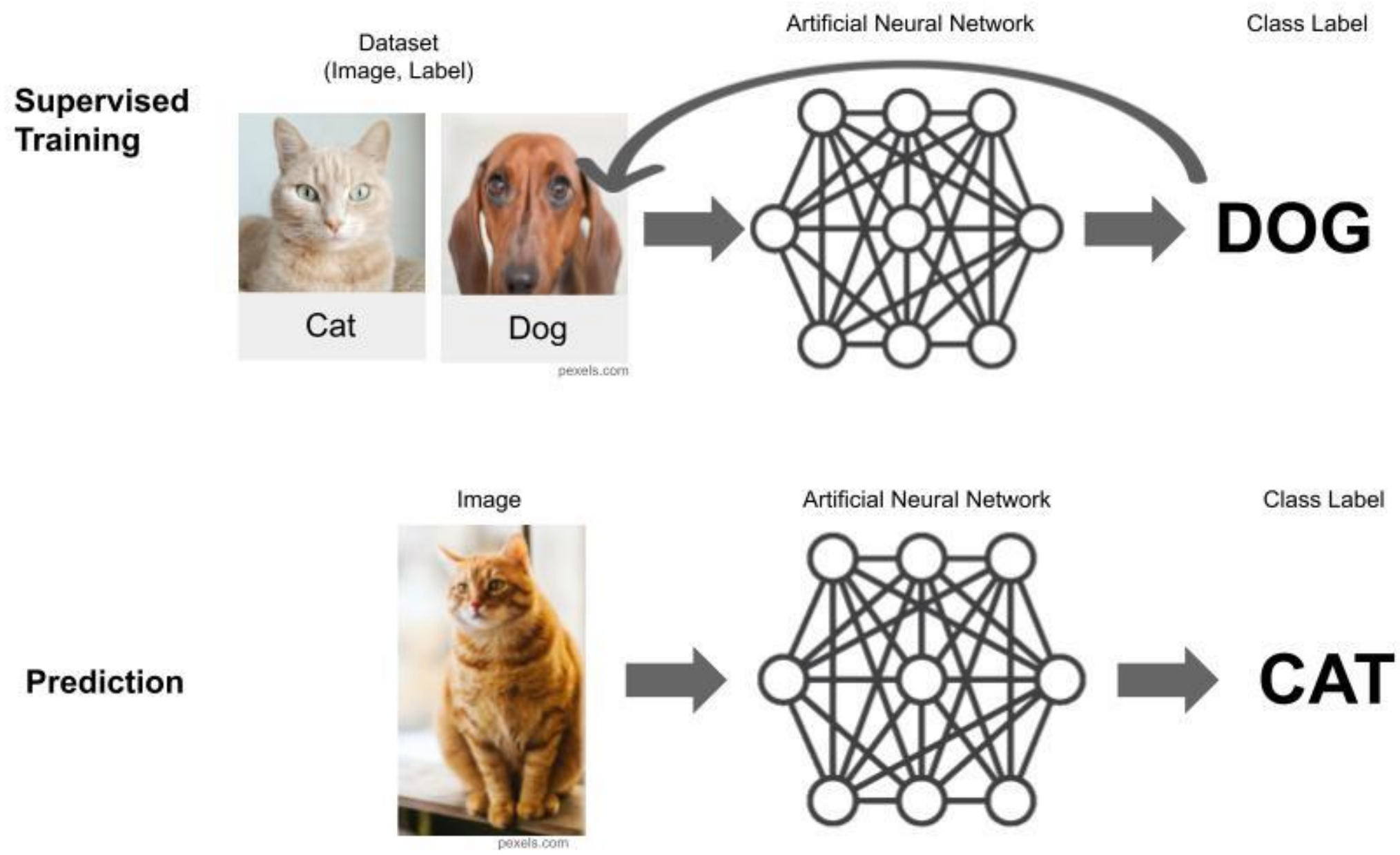

Figure 1. Image classification task powered by Deep Learning

In recent years, the state of the art of image classification has been improved by applying Deep Learning (LeCun et al., 2015). Deep learning is a type of machine learning based on artificial neural networks in which multiple complex layers of processing are used to extract progressively higher level features from data. During training, deep learning discovers intricate structure in large data sets by using the backpropagation algorithm to indicate how a machine should change its internal parameters that are used to compute the representation in each layer from the representation in the previous layer. Deep-learning methods are representation- learning methods with multiple levels of representation, obtained by composing simple but non- linear modules that each transform the representation at one level starting with the input into a representation at a higher, slightly more abstract level, suppressing irrelevant variations, in order to learn complex functions.

Currently, convolutional neural networks (CNN) are one of the most prominent approaches in deep learning (LeCun et al., 2015), including AlexNet (Krizhevsky, 2012), ResNet (He et al., 2016), MobileNet (Howard et al., 2017), EfficientNet (Tan et al., 2019) among others.

These models are typically trained with a large dataset using supervised learning. Yet, transfer learning exploits the capabilities of the pre-trained CNN, retaining both its initial architecture and all the learned weights, to new data with a smaller population instead of training a CNN from scratch.

When working on classification problems, in general, the performance of the model is measured during the validation and testing of the trained model by metrics, such as the accuracy, precision, recall among others. Accuracy is the measure of how much percentage of images are classified correctly. It is the ratio of the number of correct predictions over total predictions. Here a high value of accuracy represents better performance.

Precision measures the ability of a model to not to classify a negative instance as positive, by measuring how many positive predicted instances by the model were actually positive, while recall measures how many of the positive instances were correctly classified by the model.

Once trained, the model can be used to predict the classes of new images not seen by the model before. The result of a classification typically provides for each class in the scope of the image classification model a confidence score, a decimal number between 0 and 1, which can be interpreted as a percentage of confidence in predicting this class label. From this output of the model the result can be presented to the user in different ways, such as just showing the class with the highest confidence score, or all classes in decreasing order of their confidence scores among other ways.

Recent breakthroughs in Artificial Intelligence technologies have enabled numerous mobile applications using smaller, yet still accurate DL models enabling real-time classification of any image from the smartphone's gallery or camera as input (Martinez-Fernandez et al., 2021). As a result, there exist already a variety of image classification mobile applications in app stores focusing on diverse application domains, such as biology (e.g., plant or animal species classification), healthcare (e.g., skin cancer diagnosis), entertainment (e.g., classification of celebrities, age identification), among others. Others such as Google Lens offer to classify a wide range of objects, including but not limited to plants, animals, landmarks, and products.

Typically image classification is the main feature of such an app. Very few apps directly use the classification results as part of another functionality. The common interaction process of humans with image classification apps starts with presenting the app's feature when opening the app (Figure 2). It involves capturing an image with the smartphone's camera or uploading an image from the device gallery to the app. Some apps offer instructions on how to capture images with sufficient quality (e.g., in focus, good lighting, clean backgrounds) and/or even provide feedback on the quality of a captured image suggesting retrying, if necessary. Several apps also offer the possibility to crop or rotate the image. The user may also be able to provide additional information or context to improve the accuracy of the classification, e.g., regarding localization.

The app then uses a deployed DL model to classify the image. Once the image has been classified, the app commonly presents the results of the classification to the user directly and may provide the user with additional information, e.g., detailed information on spider species. The user may have the option to provide feedback on the classification result and/or ask for human assistance either by experts or community members. The user may also have the option to share the image or the results of the classification with others through various means, such as social media, messaging, or email.

Some apps provide information on privacy regarding the usage of user's data and images as well as information on how the DL model has been trained and its performance.

Apps focusing on the classification of objects that are potentially harmful to humans such as poisonous animals (spiders or snakes) may also present warnings with regard to taking a picture of the animal, as well as when presenting the results when poisonous species have been identified.

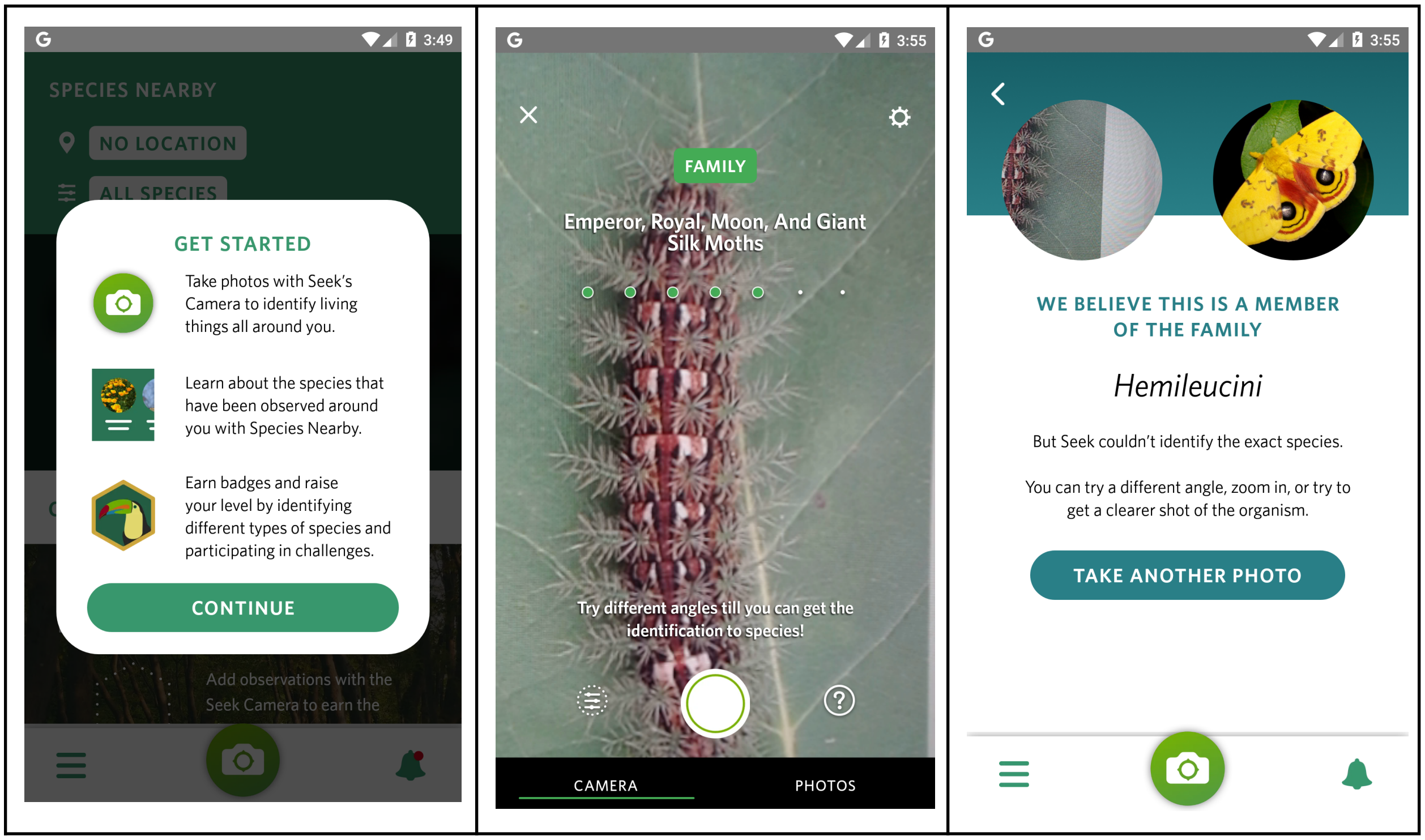

Figure 2. Example of screens of an image classification app (Seek iNaturalist)

### 2.2 Heuristic evaluation

The design of the user interface of AI-powered apps can greatly influence the user's perception of the app's functionality, accuracy, trustworthiness, and ease of use. A well-designed interface can help users understand the app's capabilities, how to interact with it, and the results they can expect. Furthermore it can help to reduce the risk of a harmful use of the results. On the other hand, a poorly designed interface can lead to confusion, frustration, and a lack of trust in the app's ability to perform as intended. Therefore, it is essential to consider the quality of the user interface for AI-powered apps to ensure a positive user experience.

In order to improve the quality of user interfaces, various methods can be used, including heuristic evaluation. Heuristic evaluation (Nielsen and Molich, 1990) is a well-known and widely accepted method used in Human-Computer Interaction to evaluate the usability of a user interface by examining it against a set of established heuristics or guidelines. Through this evaluation, designers and developers can identify potential usability issues, and prioritize and make changes to the interface design in order to improve the overall user experience. This method can be performed at any stage of the design process and is relatively quick and cost-effective compared to user testing. The results of heuristic evaluation can help identify potential usability issues early in the design process, leading to more user-friendly products.

A heuristic evaluation is based on a set of usability heuristics. Usability heuristics refer to a set of high-level guidelines or best practices, typically based on a combination of established principles from a mix of theory-based knowledge, experience and common sense. Besides the usability heuristics originally developed for GUI interfaces on desktop computers, including Nielsen's ten heuristics (Nielsen, 1994) and Shneiderman's eight golden rules (Shneiderman, 1998), various specialized heuristic sets are being developed. This includes heuristics for user interfaces on different devices, such as mobile phones (Inostroza et al., 2012)(Salazar et al., 2013), ambient displays (Mankoff et al., 2003) and virtual worlds (Rusu et al., 2011), or for specific groups of users, such as senior citizens (Salman et al., 2018), or application domains such as e-commerce (Tezza et al., 2011) or e-learning (Gu et al. 2011).

This shows that usability heuristics must be carefully selected so that they reflect the specific type of interface being designed and may require alternative heuristics or re-interpretation of existing ones in order to make sense (Holzinger, 2005). It also points out the need for adapting those "traditional" usability heuristics to fit the specific characteristics and limitations of DL-powered applications. And, although there exists first research developing usability heuristics for Artificial Intelligence systems, some are focused on specific types of tasks, such as recommendation or conversation. However, as AI/DL-powered systems vary largely with respect to their task and the ways an application may use this technology, a lack of specific heuristics to guide the design of user interfaces of mobile applications can still be observed.

## 3. Overview on existing AIX heuristics

The need for creating AIX heuristics arises from the fact that AI-powered systems present unique challenges for human-computer interaction that are not addressed by traditional usability heuristics. These challenges include issues related to explainability, transparency, trust, and ethics. Therefore, the development of a set of heuristics specifically tailored to AI systems is important to help to create more user-centered and effective interfaces that address these challenges and promote a positive user experience.

Recently, first proposals of guidance for AI design have emerged mostly in white papers by IT companies such as as Google's People + AI Guidebook (Google, 2023), Apple's Machine Learning design guidelines (2023) or Orium's smarter patterns (2023) for any kind of AI systems, while Amazon (2020), Facebook (2023) and IBM (2023) propose guidelines for conversational AI. Few academic research is available, including Dudley & Kristensson (2018) focusing on the design of interactive Machine Learning and Mohseni et al. (2021) presenting a framework for the design of explainable AI systems. Amershi et al. (2019) at Microsoft synthesized research in interaction with AI into a set of guidelines for human-AI interaction (HAI guidelines) as well as Wright et al. (2020) presenting a comparative analysis of industry human-AI interaction guidelines. These frameworks either provide general principles and guidelines or focus more on a specific type of AI-powered system (such as recommendation systems or conversational agents) that are difficult to apply to other AI tasks, such as image classification.

Considering general guidelines for creating a basis for the proposal of heuristics for image classification, Table 1 presents a mapping of the encountered guidelines.

Table 1. Mapping of general AIX guidelines

| AIX Guidelines | (Amershi et al., 2019) | (Apple, 2023) | (Google, 2023) | (Orium, 2023) |
|---|---|---|---|---|
| Make clear what the system can do | x | x | x | |
| Explain the benefit not the technology | | x | x | |
| Avoid technical jargon | | | x | x |
| Make clear how well the system can do what it can do | x | x | | x |
| Demonstrate how the users can get satisfactory results | | x | | |
| Time services based on context | x | | | |
| Show contextually relevant information | x | | x | |
| Anchor on familiarity | | | x | |
| Make the presentation of images and information accessible | | | | x |
| Show processing status | | | | x |
| Determine how to show model confidence, if at all | | x | x | x |
| Explain for understanding, not completeness | | x | x | |
| Go beyond in-the-moment explanations | | | x | |
| Match social norms | x | | | x |
| Mitigate social biases | x | | | x |
| Support efficient invocation | x | | x | |
| Support efficient dismissal | x | | | |
| Support efficient correction | x | x | x | |
| Make it clear when the results are inaccurate | | | | x |
| Provide manual controls when the AI fails | | | x | x |
| Offer customer support when the AI fails | | | x | |
| Scope services when in doubt | x | | | |
| Make clear why the system did what it did | x | x | | |
| Remember recent interactions | x | | | |
| Learn from user behavior | x | x | | |
| Update and adapt cautiously | x | x | | |
| Let users give feedback | x | x | x | |
| Convey the consequences of user actions | x | | | x |
| Provide global controls | x | x | | |
| Notify users about changes | x | x | | |
| Be transparent about privacy and data settings | | x | x | x |
| Make it safe to explore | | | x | |
| Explain how the AI model has been developed and the algorithm | | | | x |
| Give risk alerts especially in critical applications | | | | x |

Most of these AIX guidelines are presented as a list of principles or heuristics decomposed into a set of items to be used during the design of user interfaces for AI-powered systems. No guidelines in the form of a heuristic evaluation checklist supporting also the evaluation of such interfaces were encountered. There also does not exist any (automated) tool support for such an evaluation of this kind of interface. In this way the results demonstrate the current lack of a set of AIX guidelines tailored specifically for applications with image classification.

## 4. Research methodology

We developed the AIX guidelines for image classification apps using a systematic methodology proposed by Quiñones et al. (2018) and Rusu et al. (2011):

Step 1. At this exploratory stage we reviewed literature related to AIX principles and guidelines, their characteristics, as well as usability heuristics for this specific kind of application.

Step 2. At this experimental stage we analyzed the specific characteristics and usability problems based on an analysis of 100 randomly selected mobile applications with image classification available at Google Play.
Step 2. At this descriptive stage we mapped the UAIX guidelines we encountered in order to highlight the most important items of the previously collected information.
Step 3. At this correlational stage we identified a set of heuristics and selected items that mobile applications with image classification should adhere to based on the mapping of existing general AIX guidelines and the usability problems observed during the analysis of the apps.
Step 4. At the specification stage we formally specified the set of usability heuristics. We also decomposed the heuristics into a set of checklist items to facilitate their application as part of a heuristic evaluation. In addition we defined a response scale for the checklist. We described each item by an ID, question and brief explanation. For each of the items we also present an example of the item's compliance and/or violation.
Step 5. In addition we prepared an online course demonstrating and explaining the AIX heuristics and the checklist as well as a web-based tool to support the execution of the heuristic evaluation using the checklist.

## 5. Initial proposal of AIX heuristics and checklist

Based on the literature review and the experimental results testing mobile applications with image classification we defined a set of heuristics and checklist as presented in Table 2. These heuristics and items are specifically related to the image classification functionality. For a comprehensive evaluation they should be completed by general usability heuristics as well as usability heuristics customized for mobile applications.

Table 2. AIX - image classification heuristics and checklist

| **Heuristic** | **Checklist item** | **Item explanation** | **Example** | **Yes, No, N/A (*Not Applicable)*** |
|---|---|---|---|---|
| **Make expectations and limitations explicit** | 1. Does the app make it clear what kind of object it can classify? | The app presents the classes that it is able to distinguish (e.g., plants, dogs breeds) before the user can classify an image (e.g., on the home screen). | 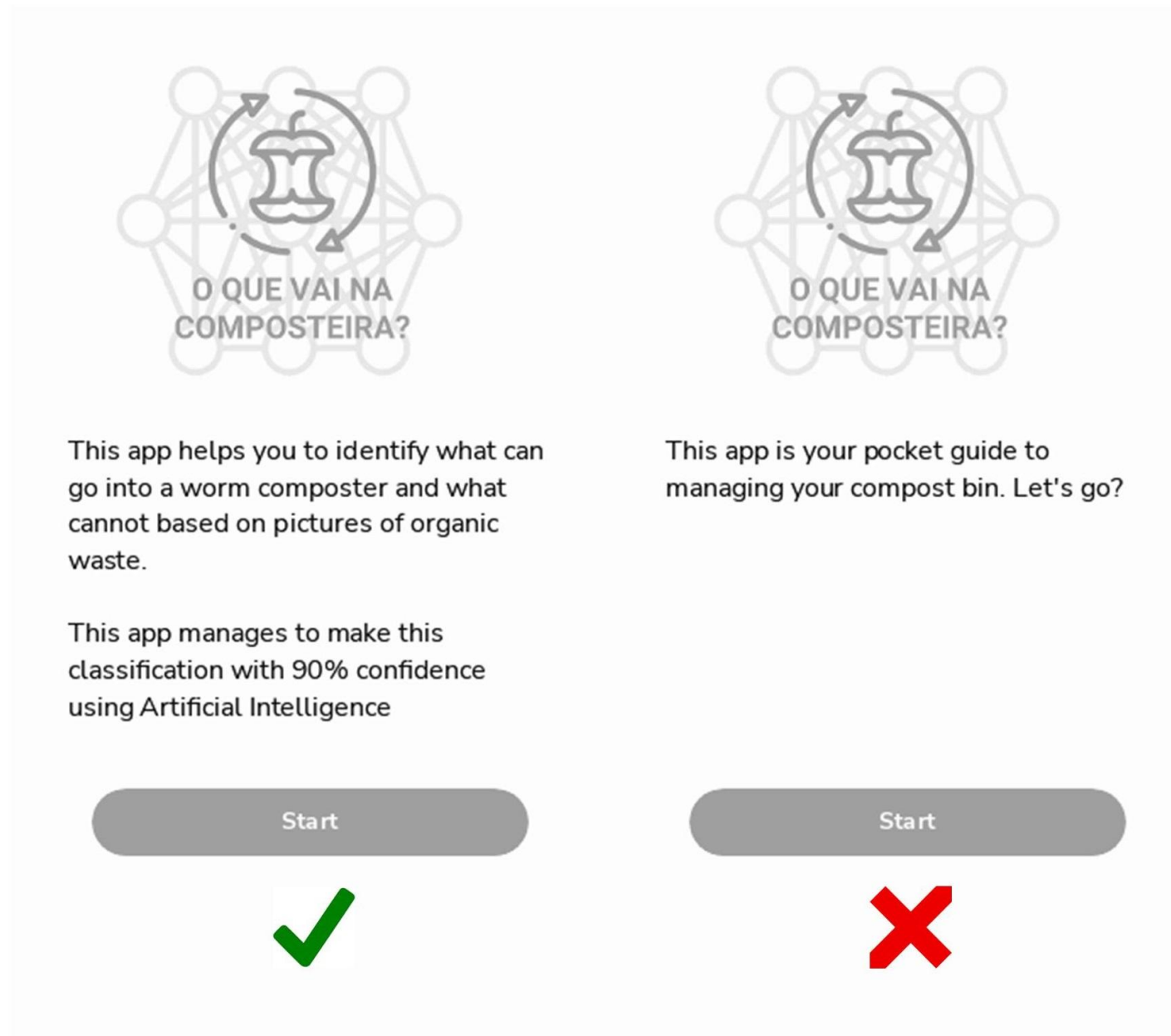 | Yes/No |
| | 2. Does the app explain which classes it can classify? | The app lists all classes it is able to classify. | 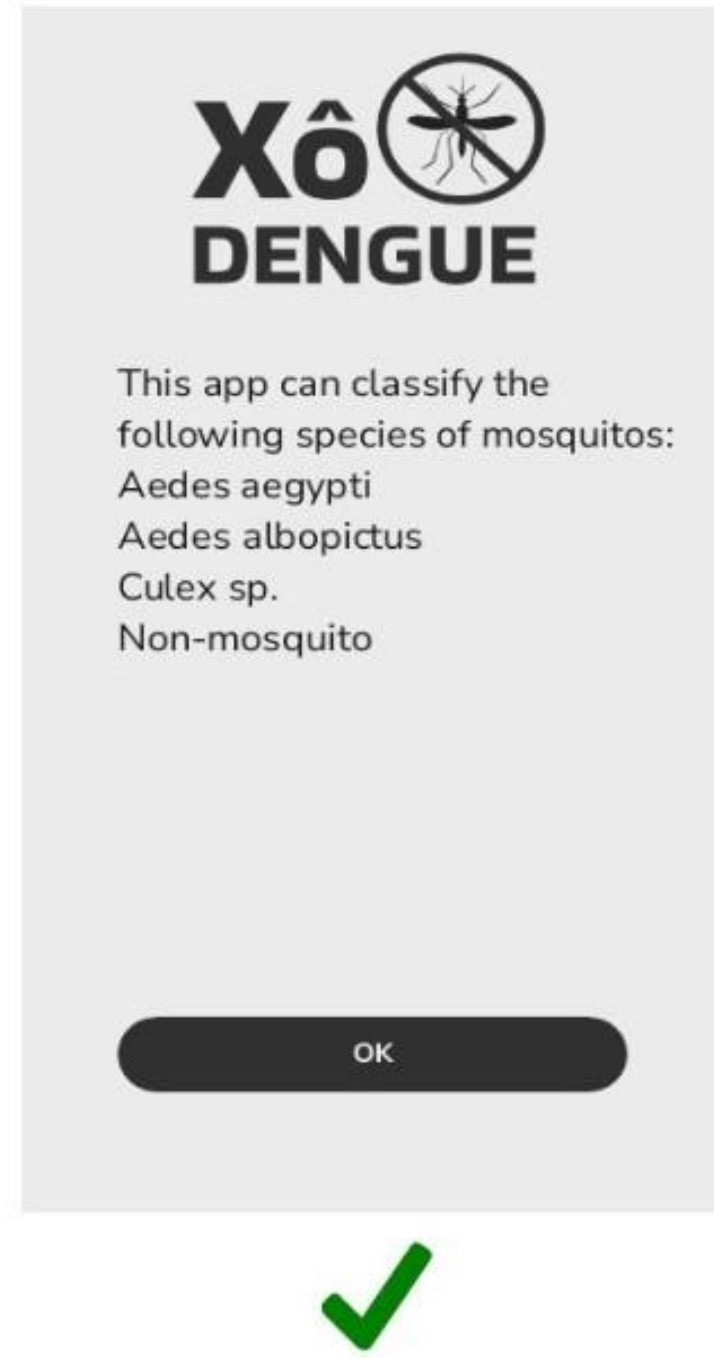 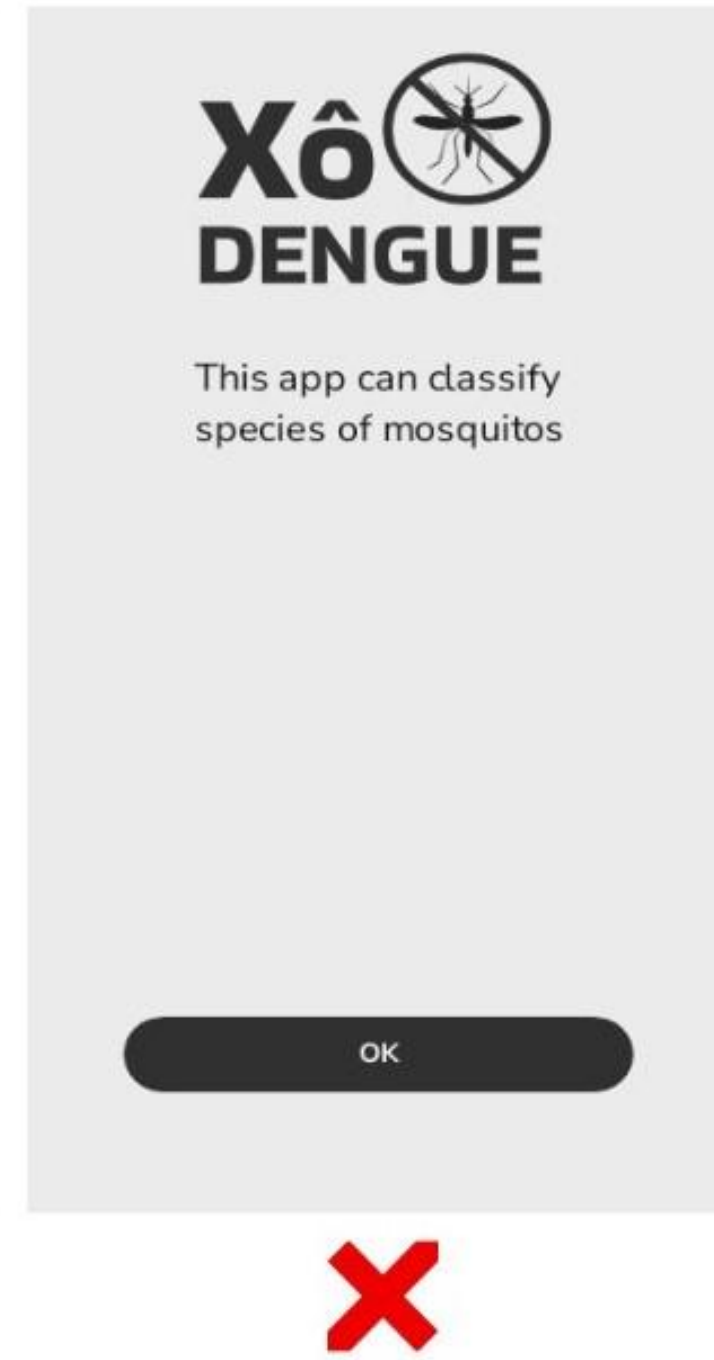 | Yes/No |
| | 3. Does the app make it clear how well it can do the image classification? | The app informs the user about the degree of its performance (e.g. accuracy) before the user can classify an image. | 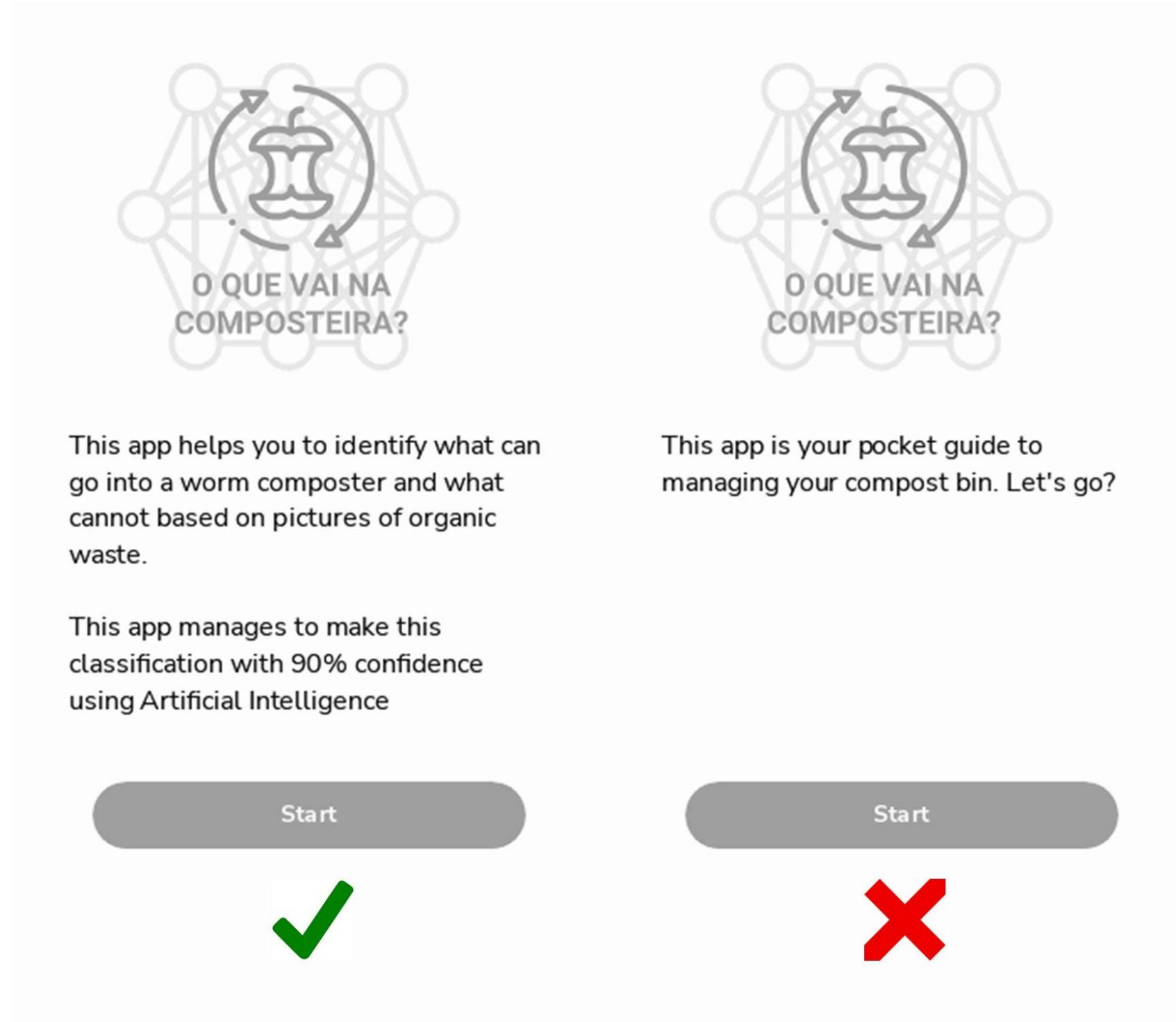 | Yes/No |
| | 4. Does the app provide understandable explanations? | The app only uses terminology understandable by the target audience, avoiding technical jargon, when presenting expectations and limitations. | 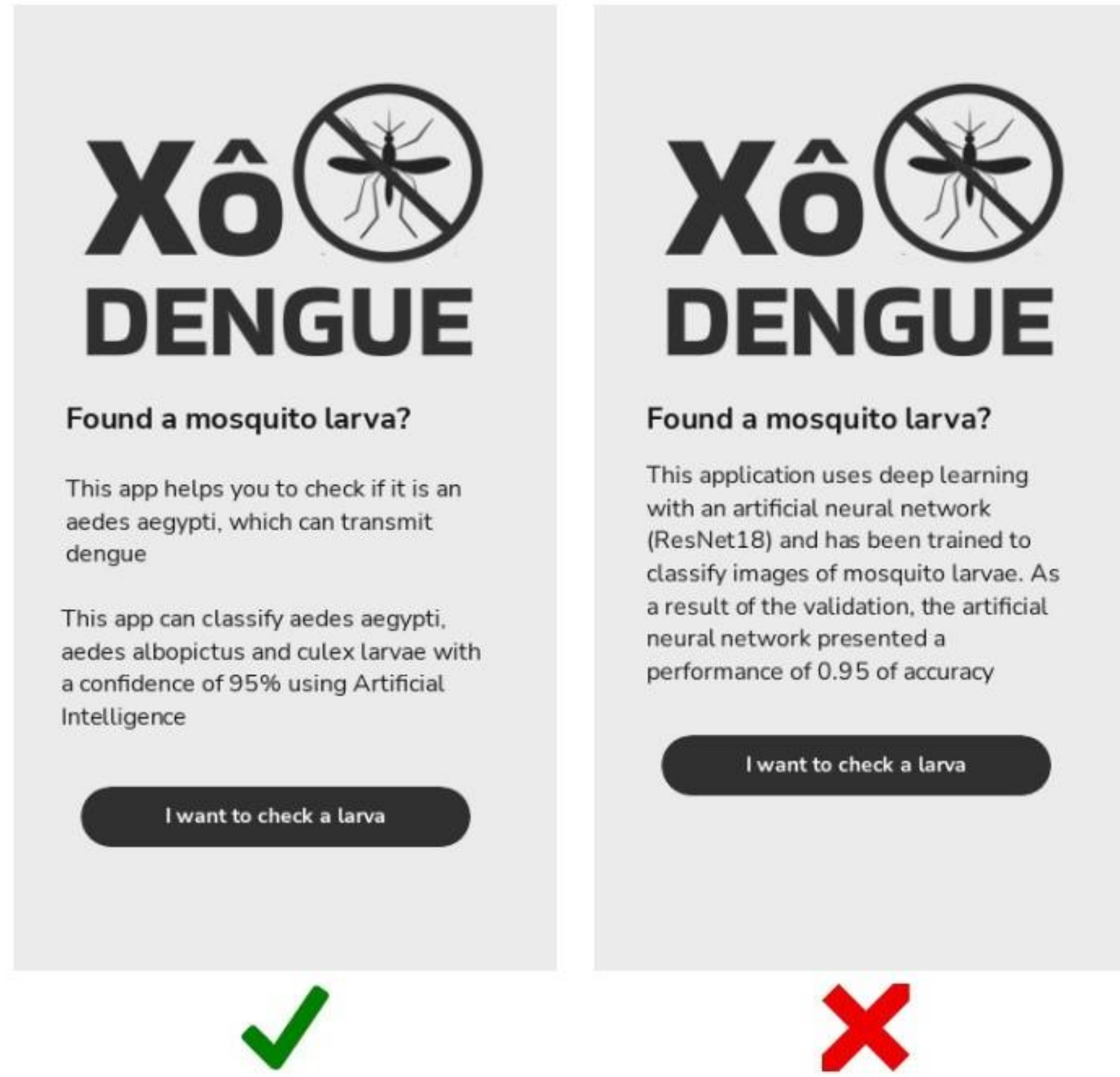 | Yes/No |
| **Support the effective use** | 5. Does the app show instructions on how to take pictures with adequate quality? | The app presents instructions/tips to guide the user to take pictures with adequate quality for classification. | 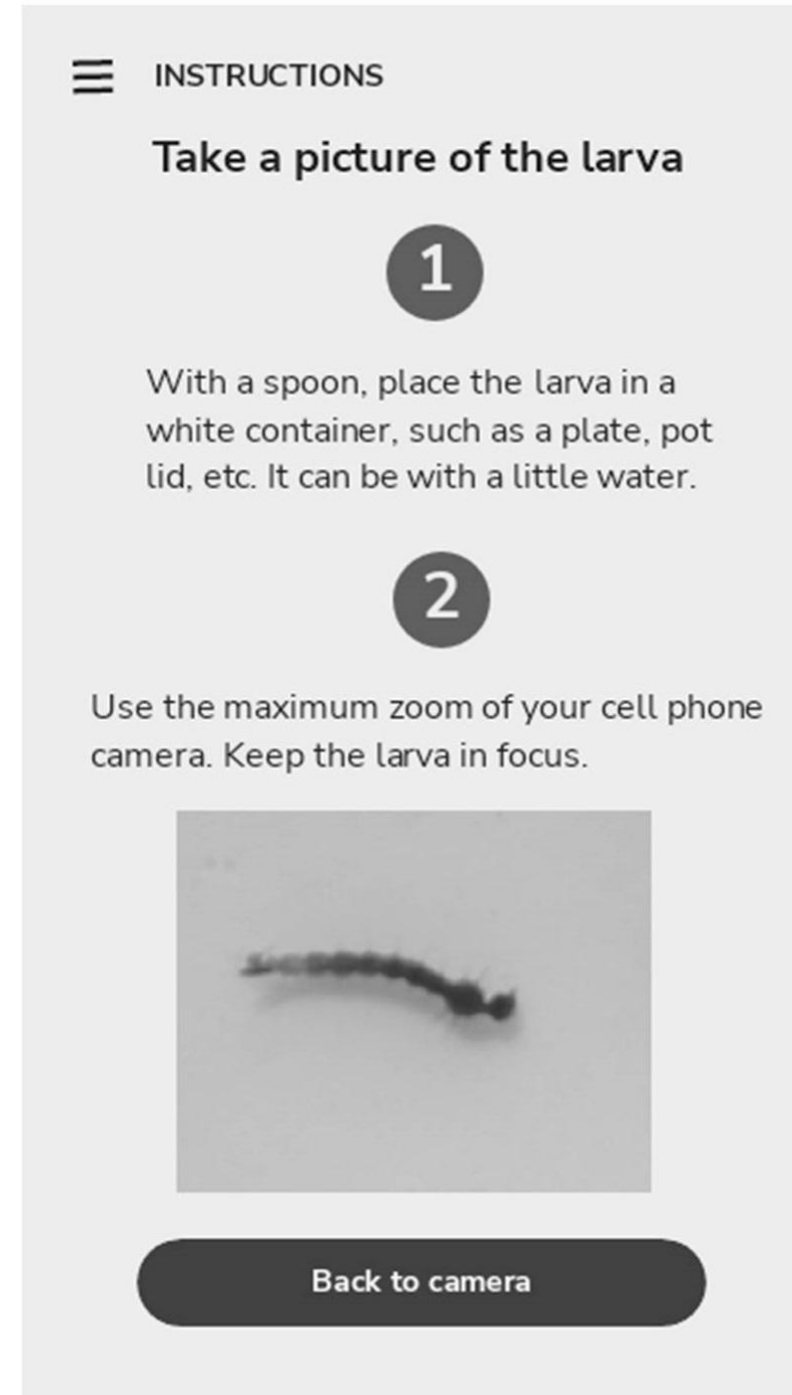 | Yes/No |
| | 6. Does the app visualize the status while processing the classification? | The app presents a progress bar to visualize the processing during classification. | 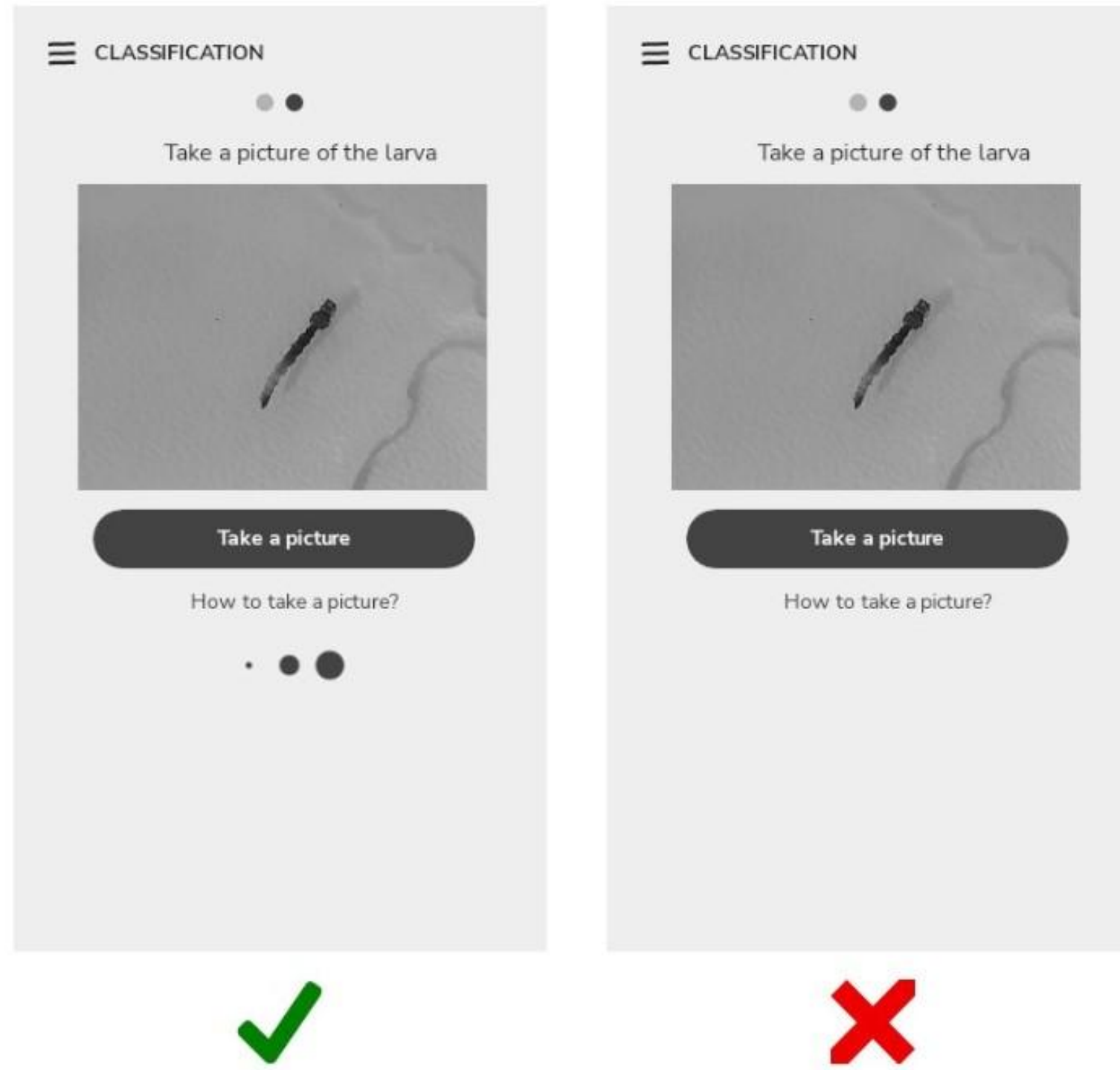 | Yes/No |
| **Support user understanding of uncertainty and model confidence** | 7. Does the app make it clear that there is uncertainty regarding the outcome of the classification? | The result of the classification is presented in a way that makes clear that there is uncertainty in relation to this result. | 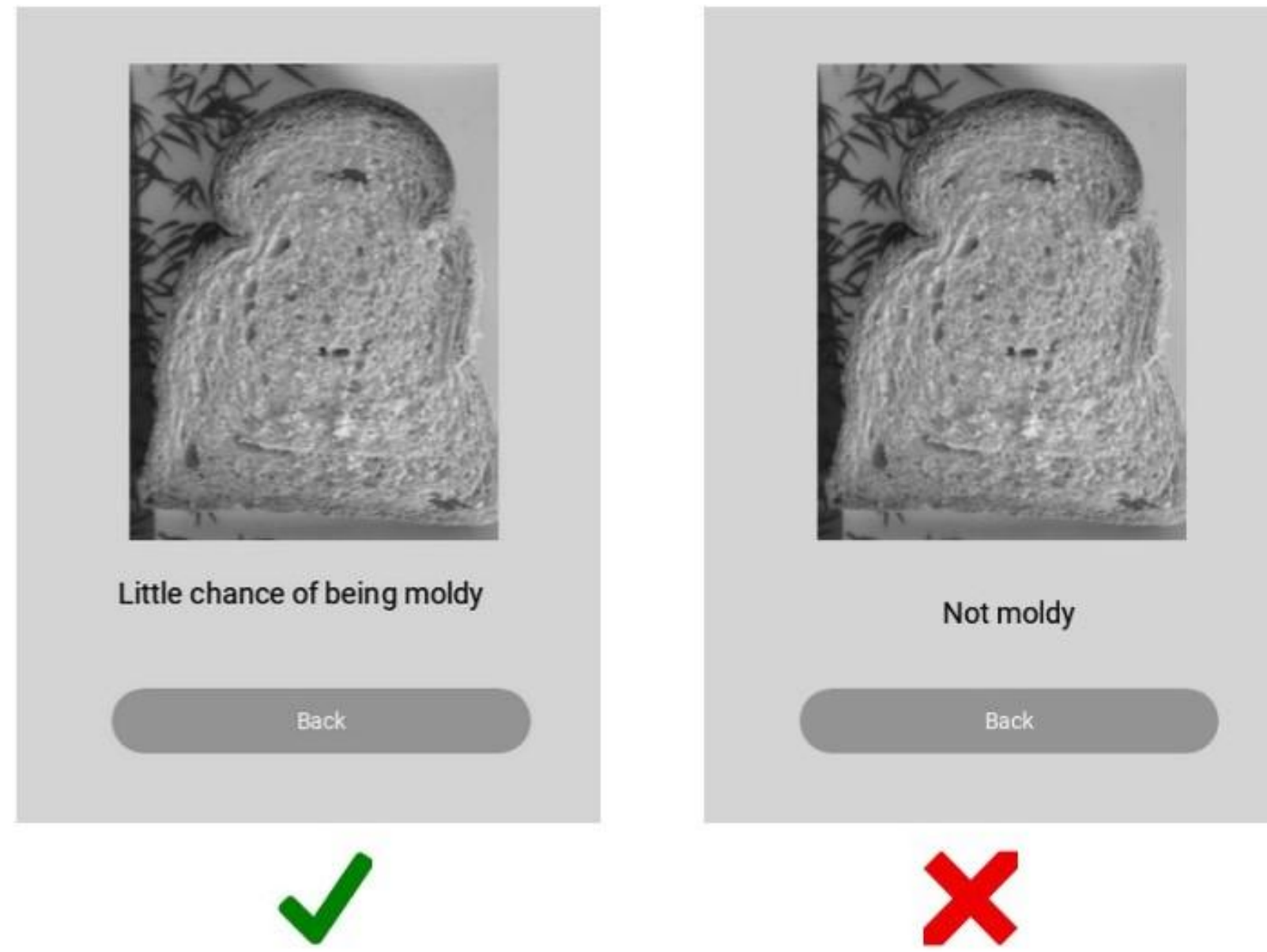  | Yes/No |
| | 8. Does the app indicate uncertainty in a way that is understandable by the target audience? | The classification result is presented in an understandable way for the target user, e.g. using categorical values such as high/medium/low or very likely/probable/not likely or presenting the n-best answer alternatives. The presentation of only confidence percentages is avoided.<br>NA: if the app does not indicate uncertainty with the classification result | Considering any citizen as a target audience in this app, the use of percentages that may not be understood by the entire target audience should be avoided<br>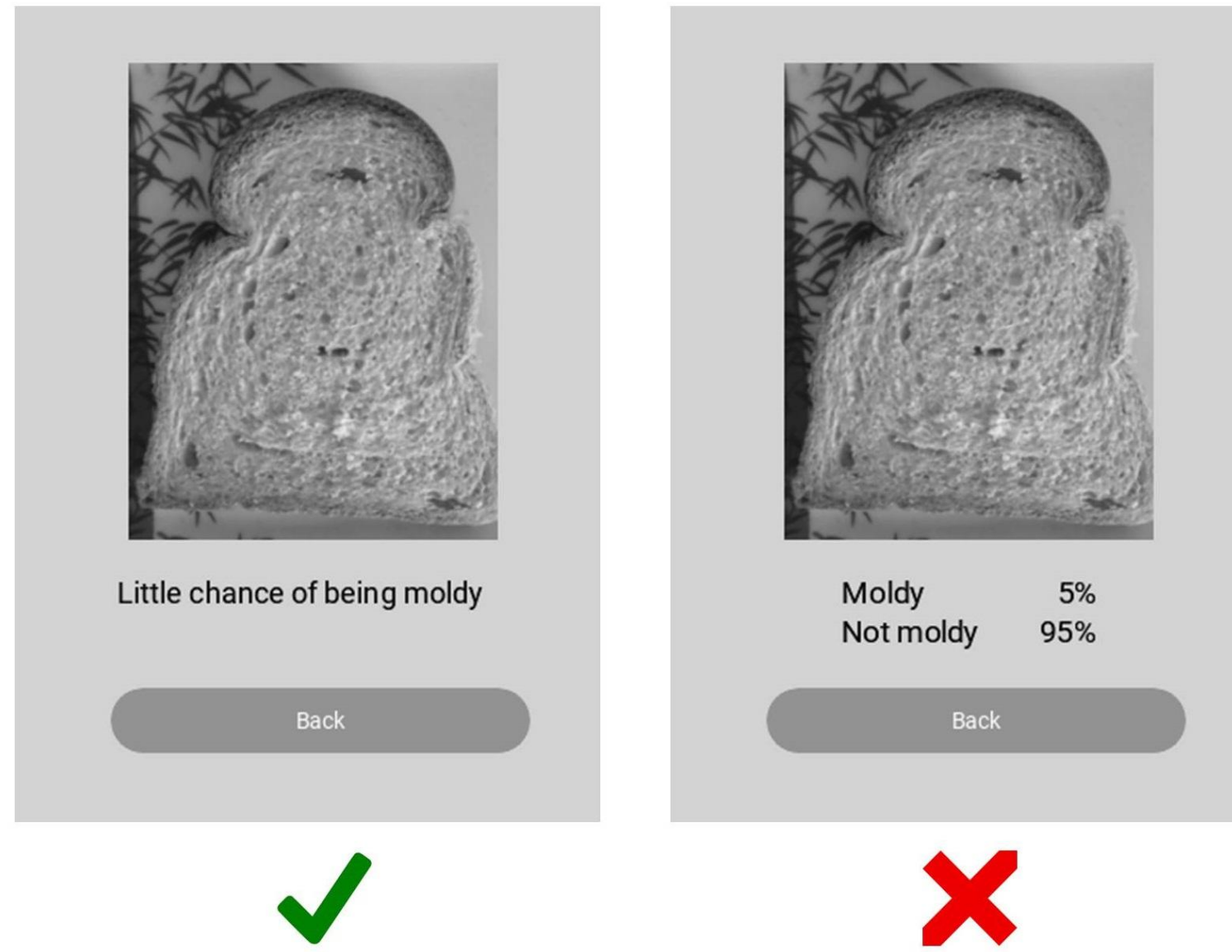 | Yes/No/NA |

<table>
<tr><td></td><td>9. Does the app demonstrate the results in a useful way?</td><td>The result of the classification is being presented in a way that helps the user to make a decision according to the use case (e.g. when classifying venomous spiders it also indicates what kind of medical assistance should be sought). NA: if the sole purpose is only the indication of a class label</td><td>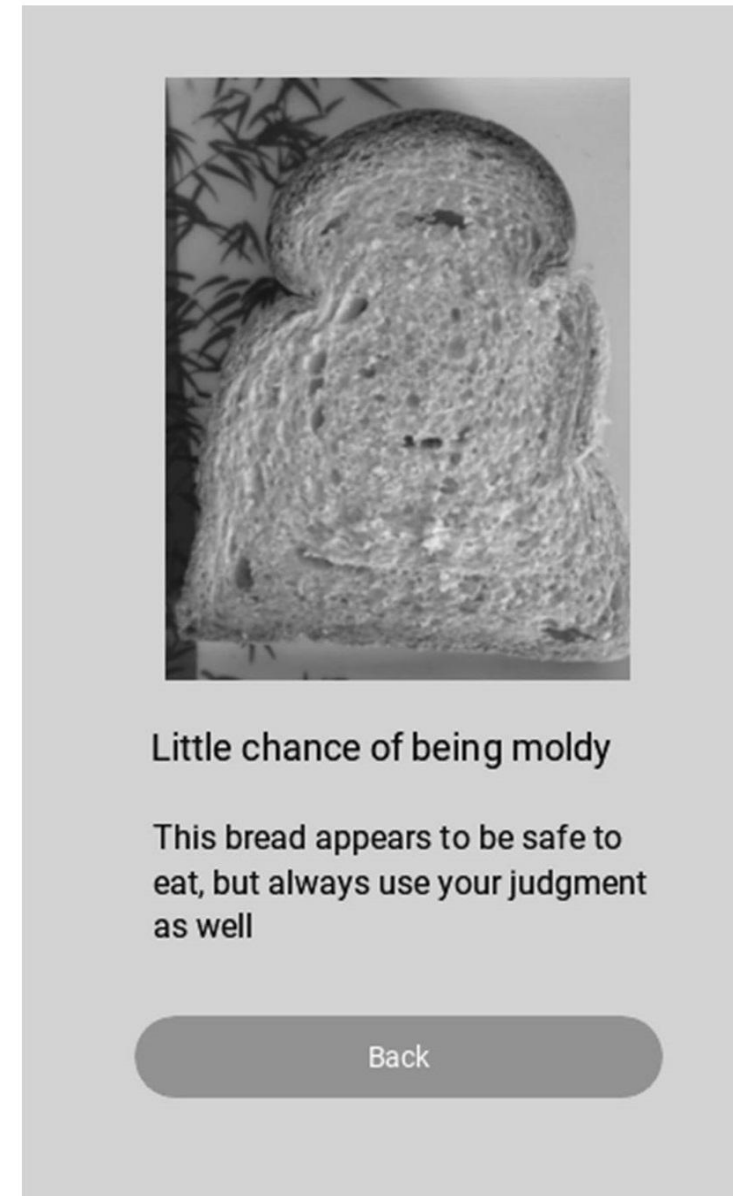

</td><td>Yes/No/NA</td></tr>
</table>

| | | | | |
|---|---|---|---|---|
| | 10. Does the app make it clear that there is uncertainty when using the classification result as part of another feature? | The app introduces the uncertainty of the classification result in some way, even when it is used directly as part of another functionality (e.g., marking locations on a map that provide the classified object).<br>NA: Image classification is not integrated into other functionality | 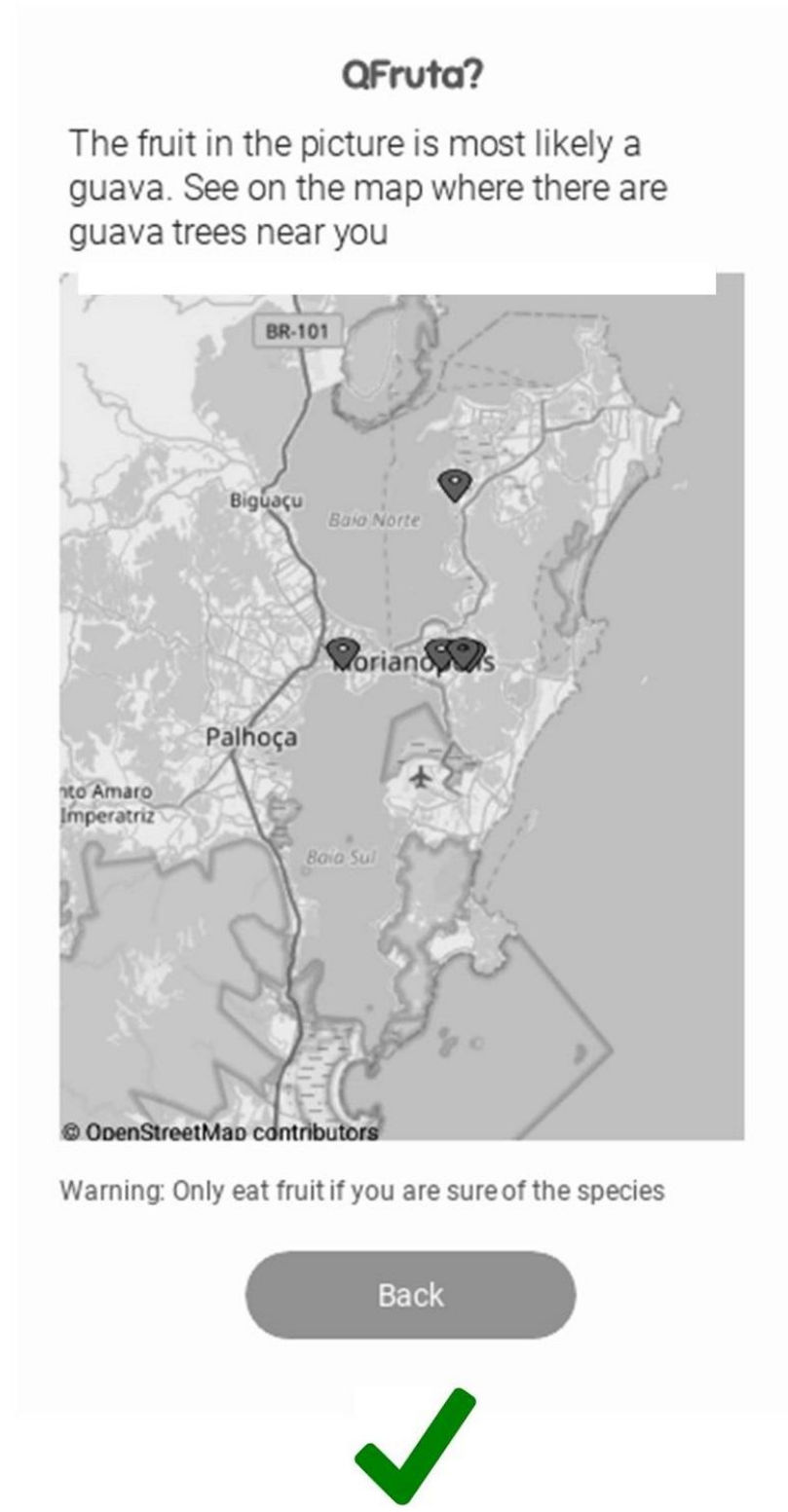<br><br>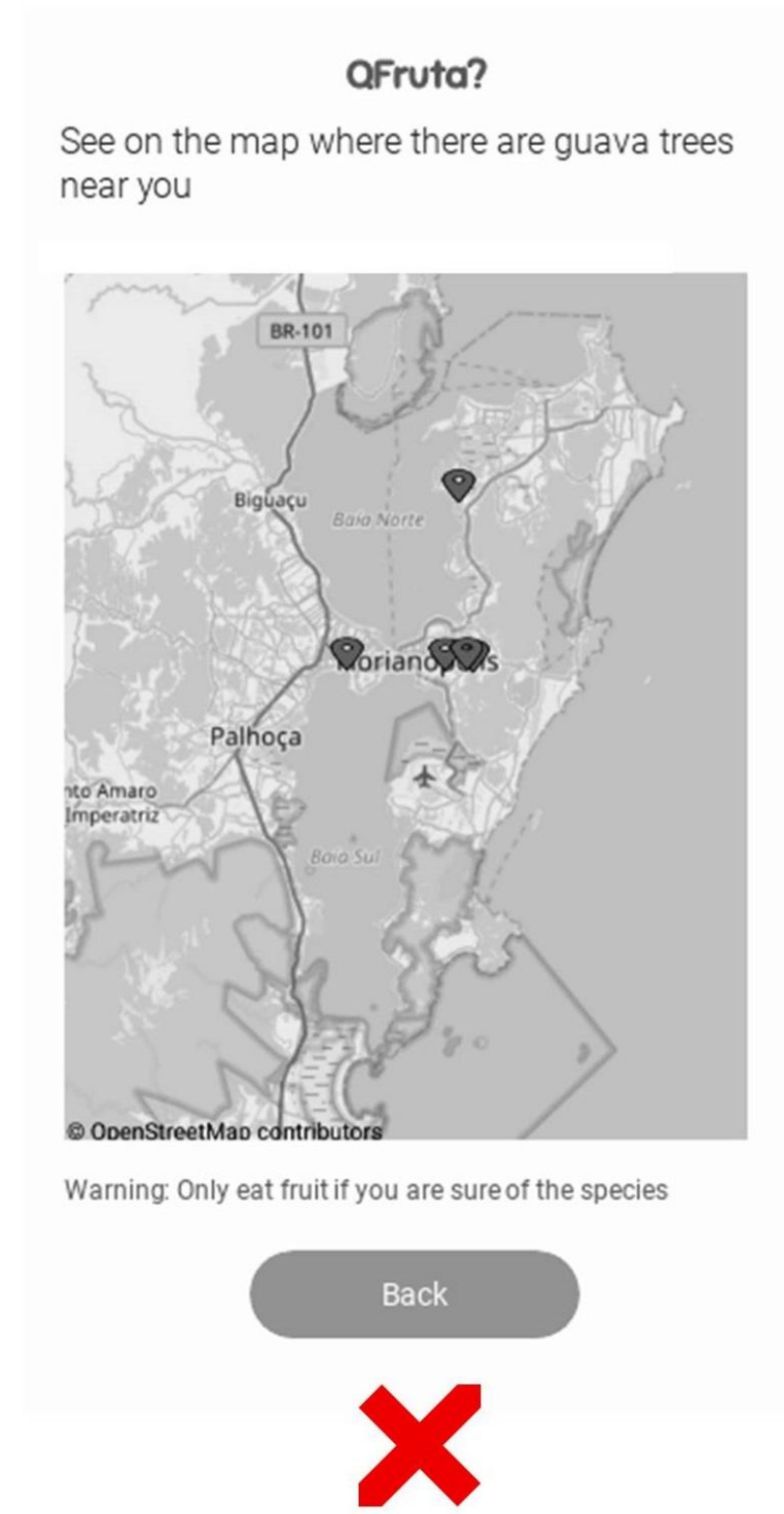<br> | Yes/No/NA |
| | 11. Does the app provide information about how the ML model was developed? | The app shows information about the development of the ML model, including eg. information about the amount of images used in training, by whom the images were labeled, what type of ML model is being used and its performance. | 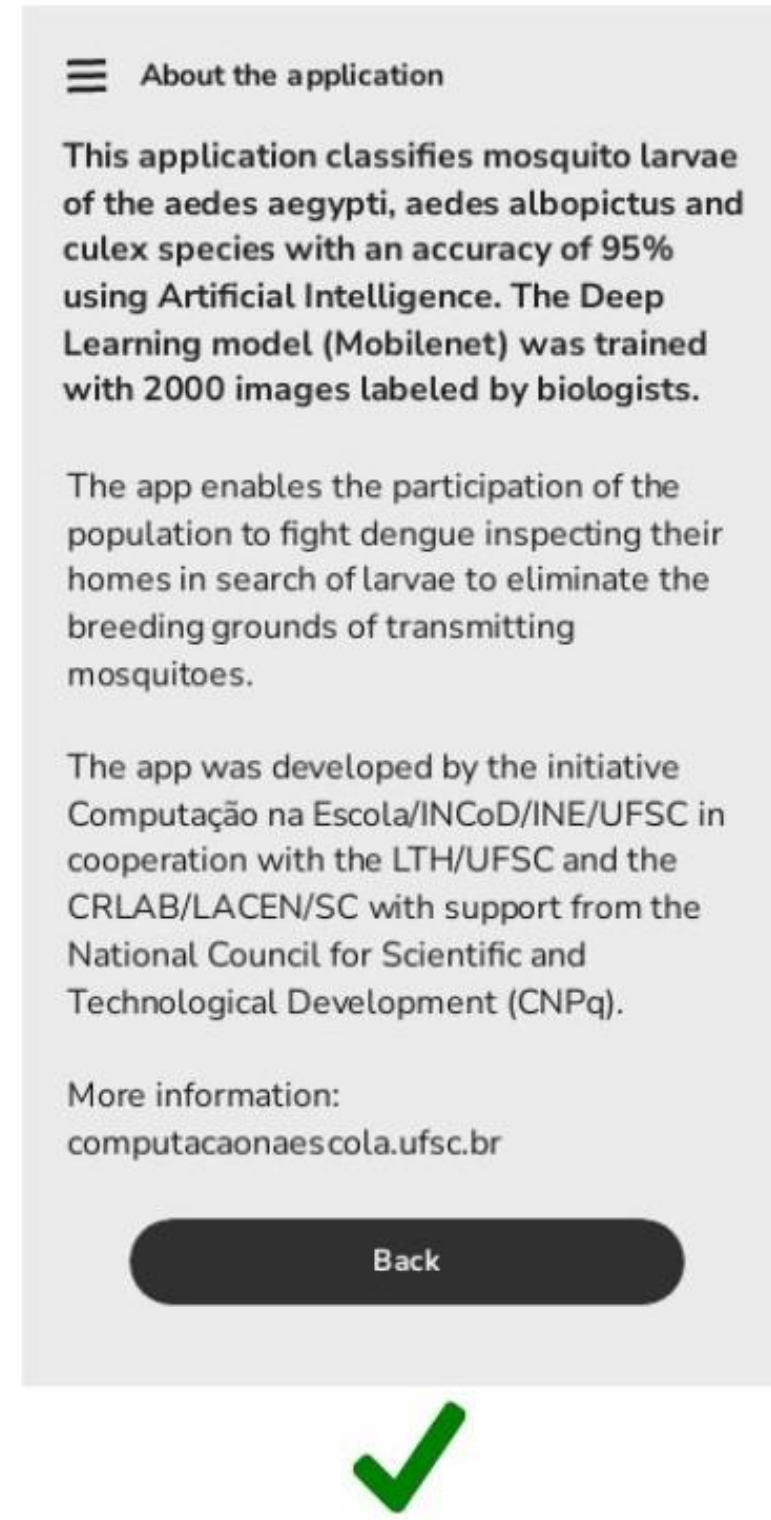 | Yes/No |
| **Ensure data privacy and security** | 12. Does the app provide information about the use of user's photos being classified? | There is information about how the user photos to be classified are used and if they are (or are not) persistently stored. If they are stored, the conditions of this storage and access to these images is specified. | 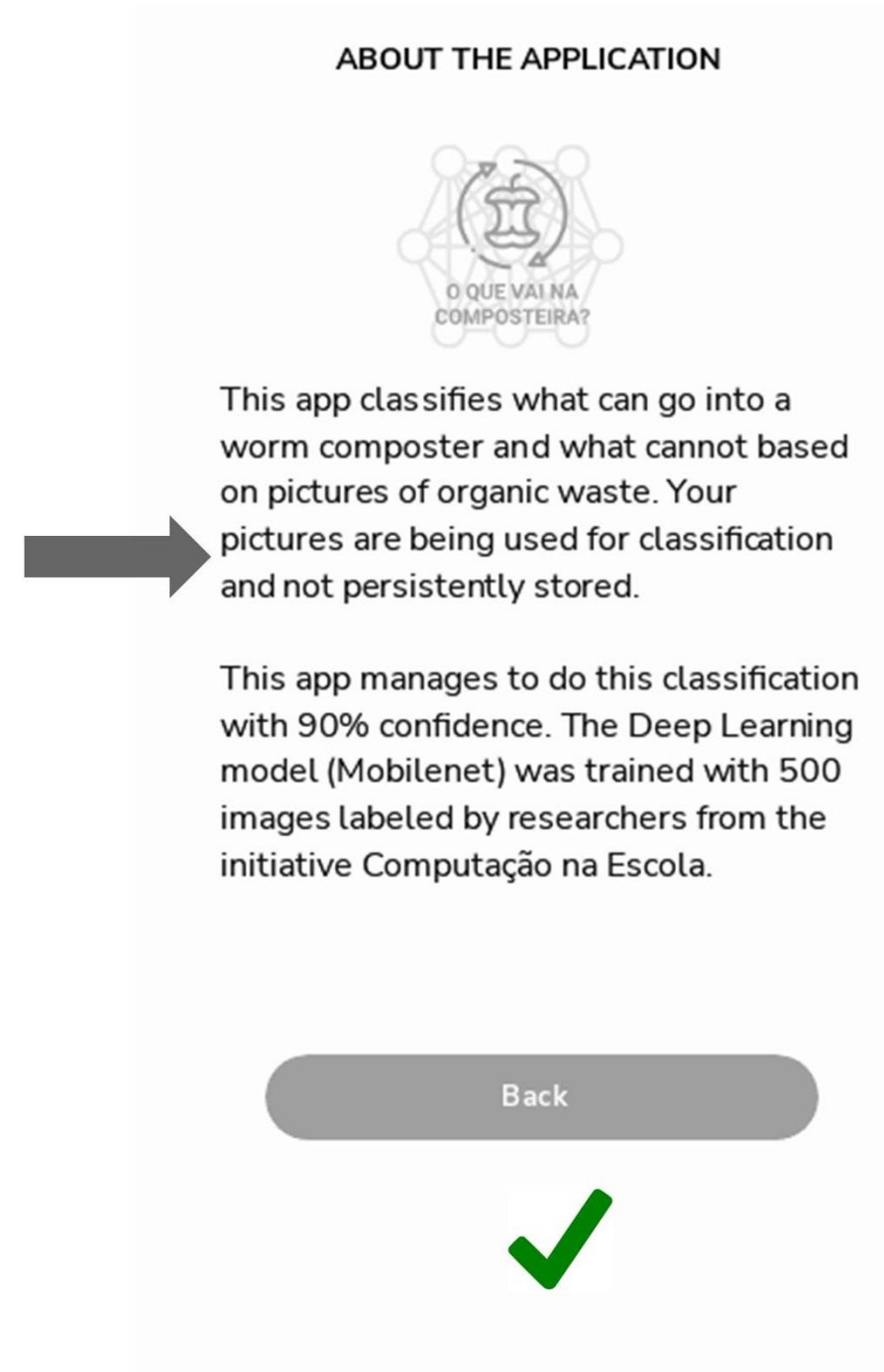 | Yes/No |
| **Fail gracefully and support error recovery** | 13. Does the app support error recovery? | The app allows the user to easily redo the classification when a classification error occurs (e.g. keeping the button to take a picture). | 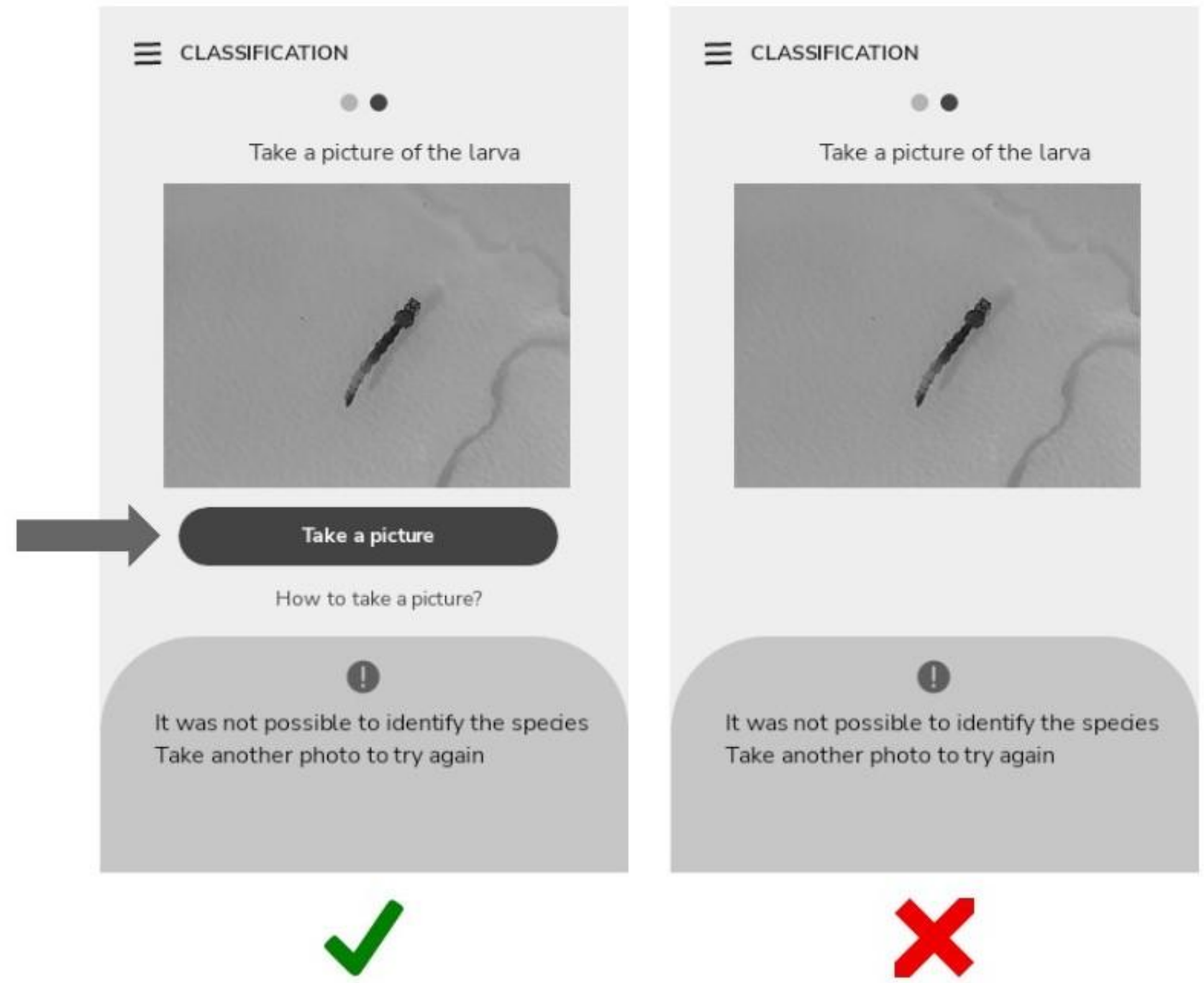  | Yes/No |
| | 14. Does the app help the user to recover from possible errors? | The app explains what to do when a classification error occurs (e.g. asking the user to take another photo). | 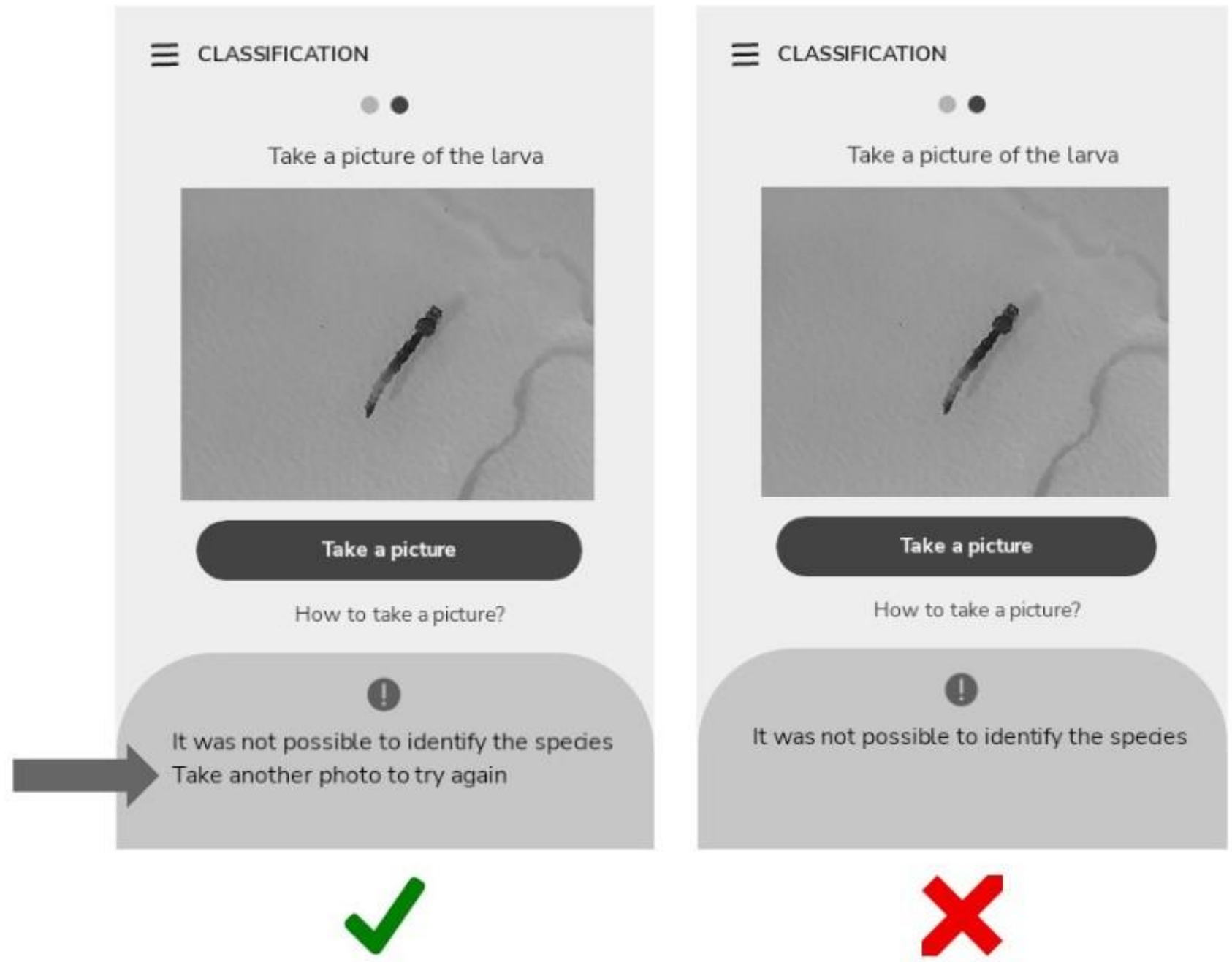 | Yes/No |
| | 15.Does the app indicate when the user tries to classify objects outside of its scope? | If the image is of an object outside the scope of the app (e.g. a glass in a dog classification app), the app displays as a result the information that this is an object outside the scope of this app's classification.<br>NA: if the app aims to classify any object | 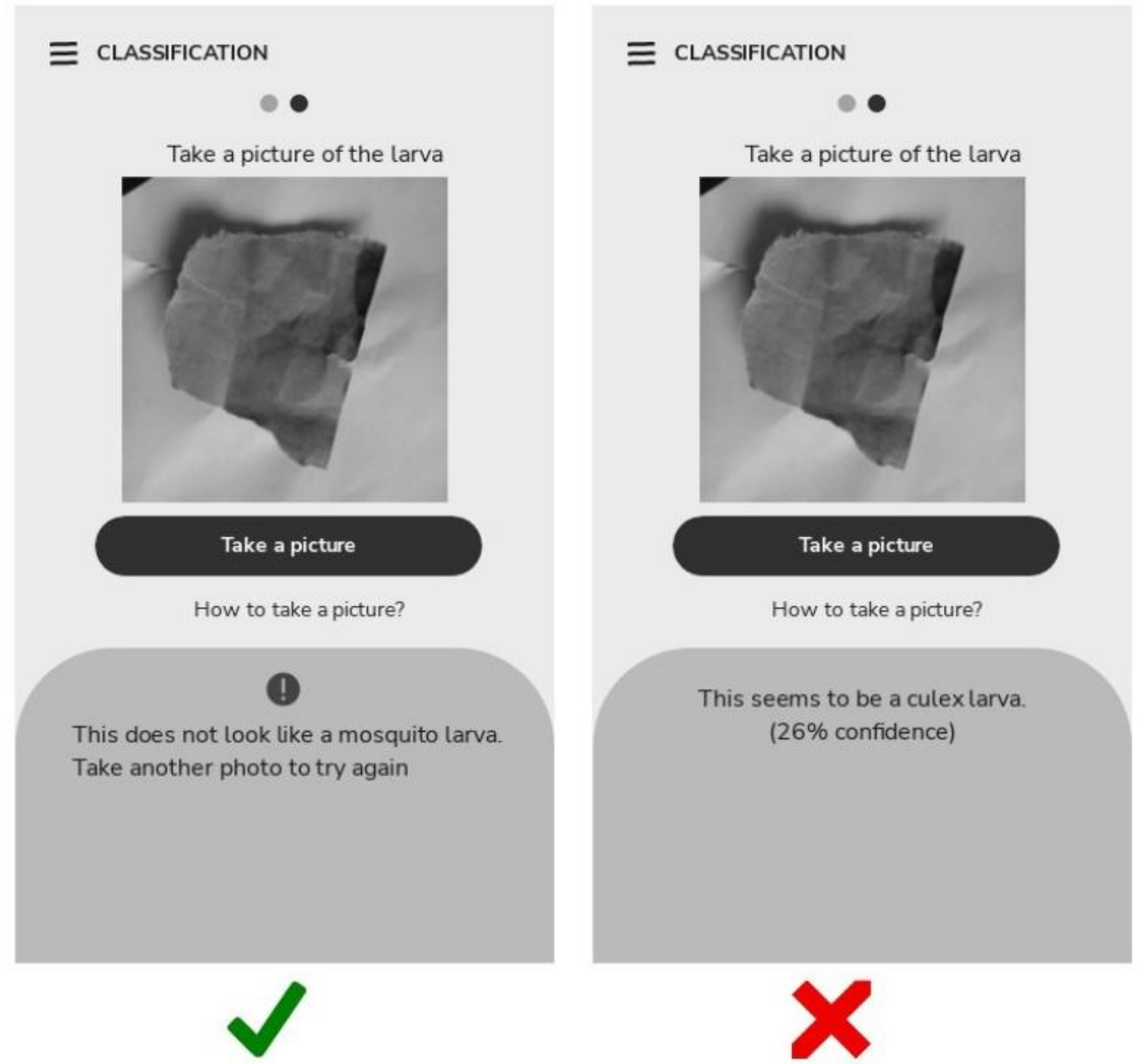 | Yes/No/NA |
| | 16. Does the app show a warning when the system is not able to classify the photo with a minimum level of confidence? | In cases where it is not possible to classify the photo with sufficient confidence (e.g. < 70%), the app alerts the user informing that it was unable to classify this image (instead of showing results even with a low confidence level). | 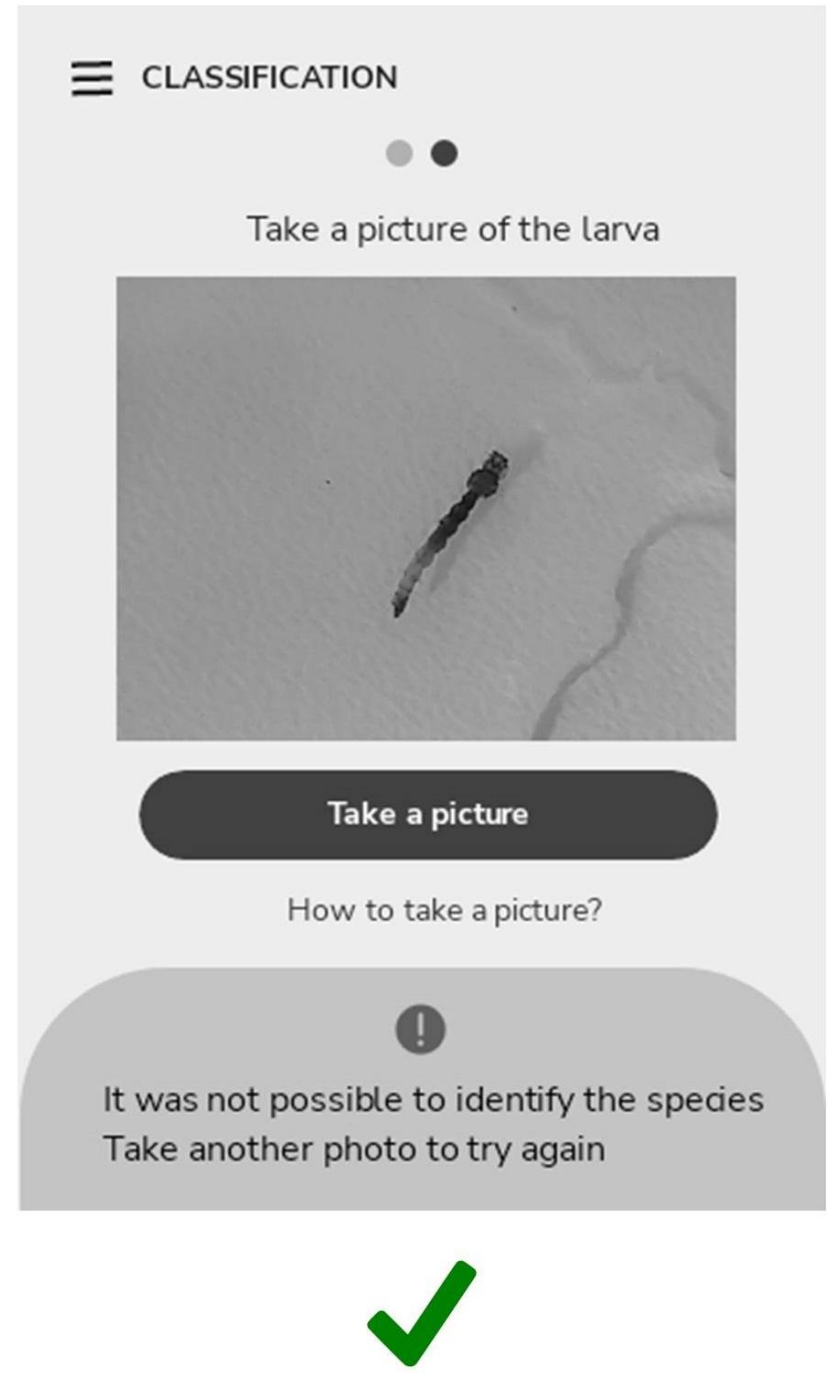  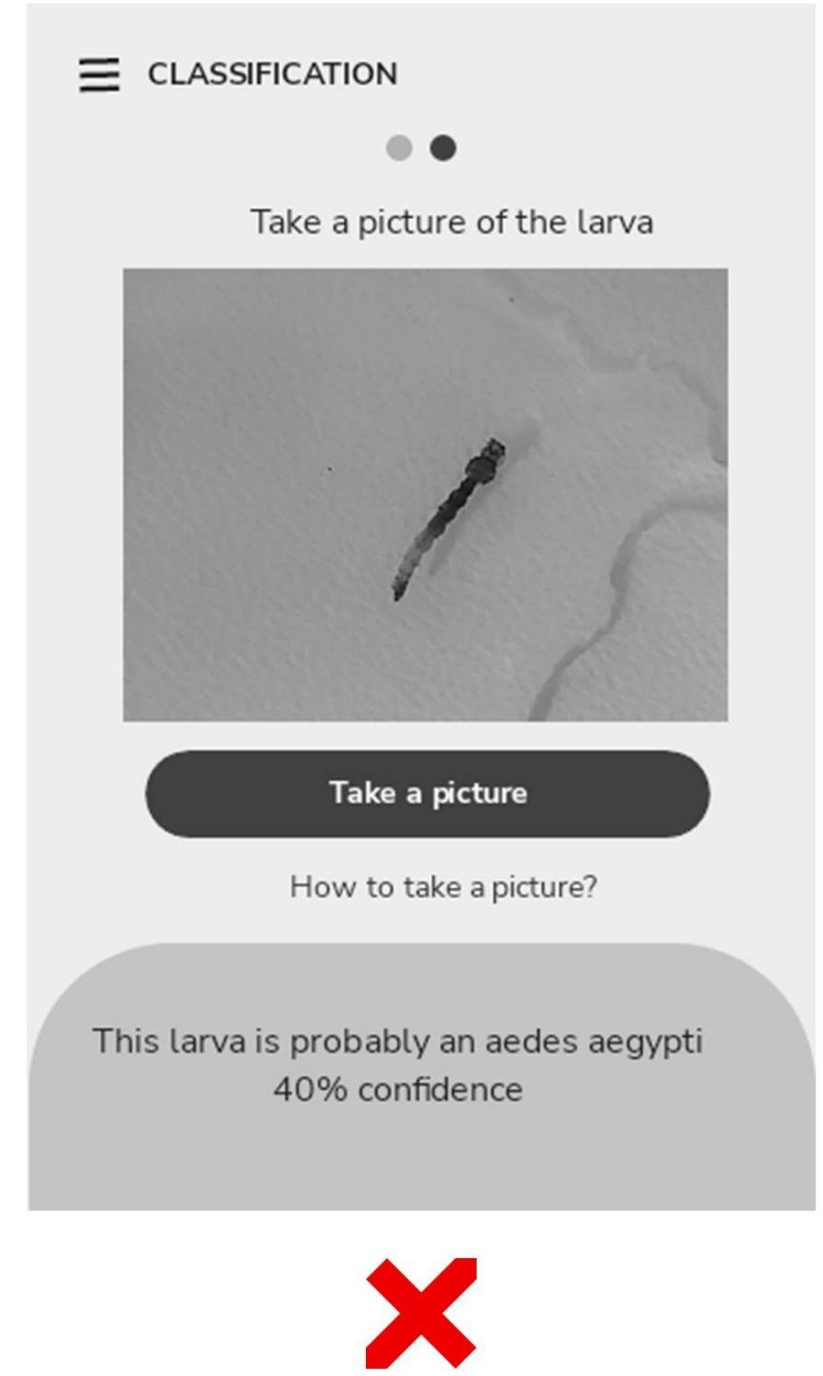  | Yes/No |
| | 17. Does the app allow the user to ask for a verification of the results by human experts? | The app provides a means of contacting domain experts to verify the classification result.<br>NA: If the app does not need the safety of human validation. | 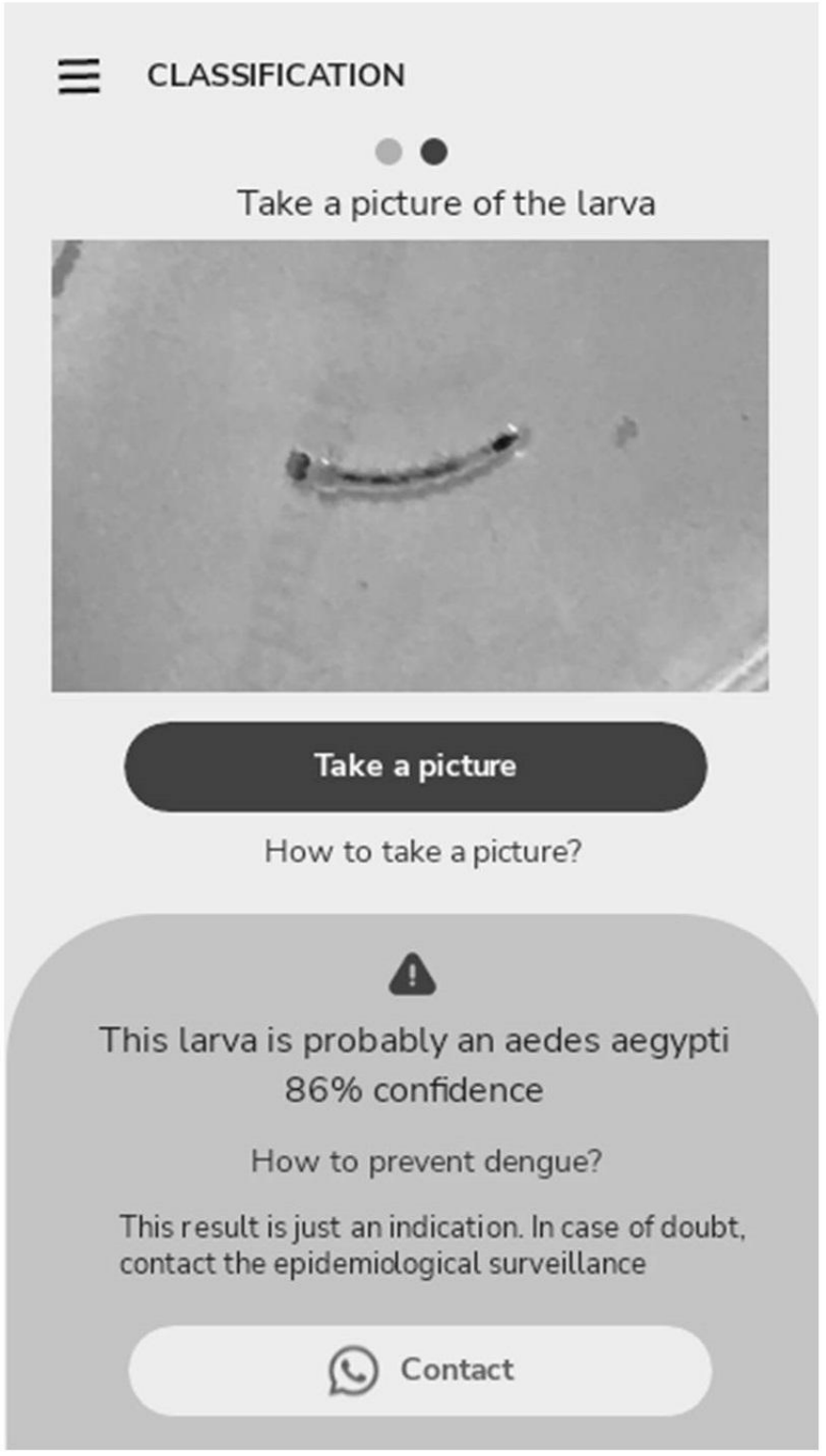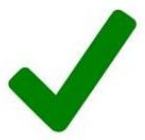 | Yes/No/NA |
| **Enable collection of user *feedback*** | 18. The app allows users with knowledge in the application domain to send feedback regarding the classification result? | The app makes it possible for a user with domain knowledge to indicate whether a classification result is correct or not.<br>NA: if the app targets a target audience without domain knowledge | App for classifying fungi for biologists in the field of mycology 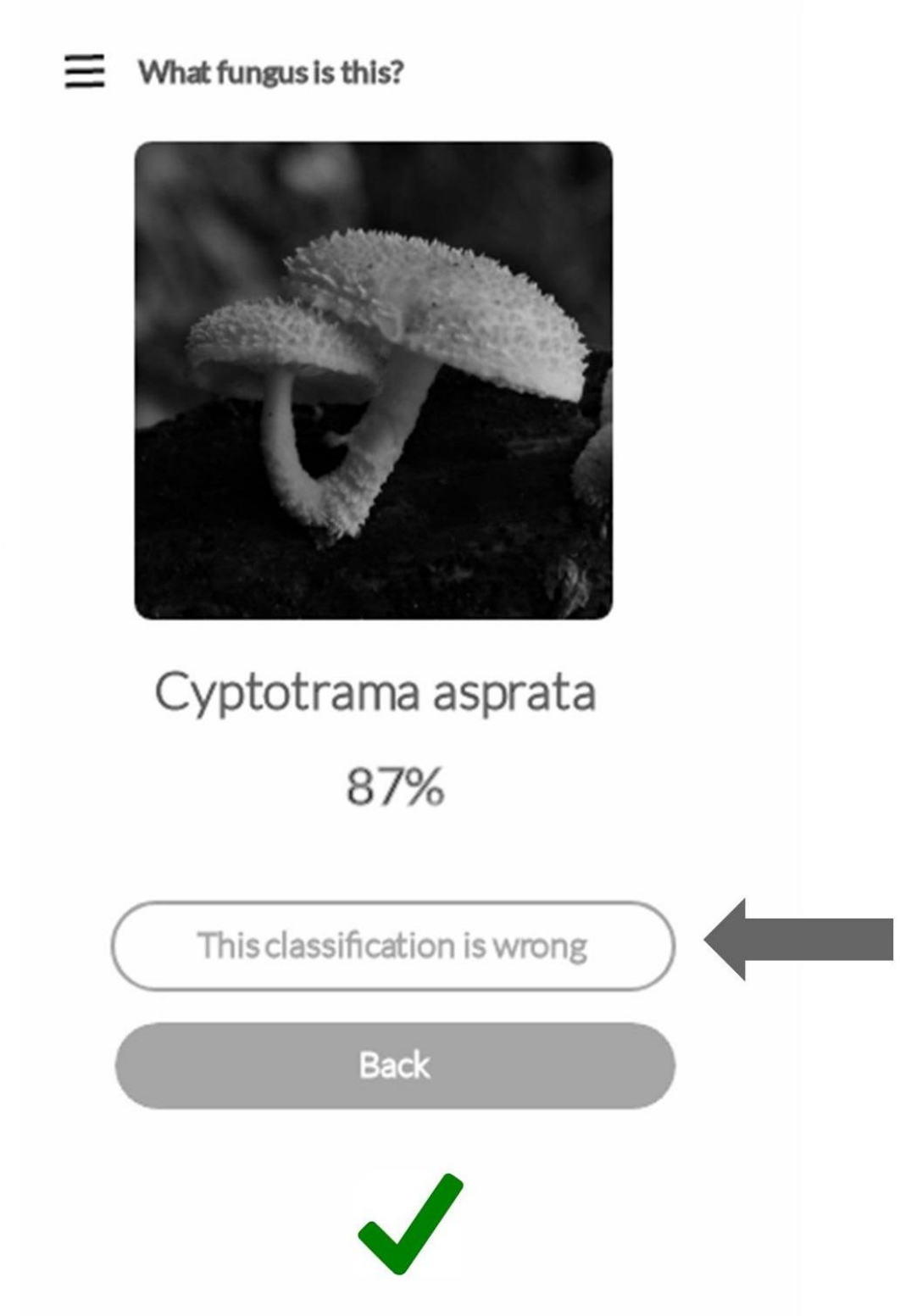  | Yes/No/NA |
| | 19. The app prohibits users without knowledge in the application domain to send feedback regarding the classification result? | The app does not offer the option to send *feedback* on the correctness of classification results to users who might not provide such feedback correctly.<br>NA: if the app targets a target audience with domain knowledge | App for classification of mosquito larvae intended for the general public without specific knowledge on the subject<br>Without the possibility to send feedback<br>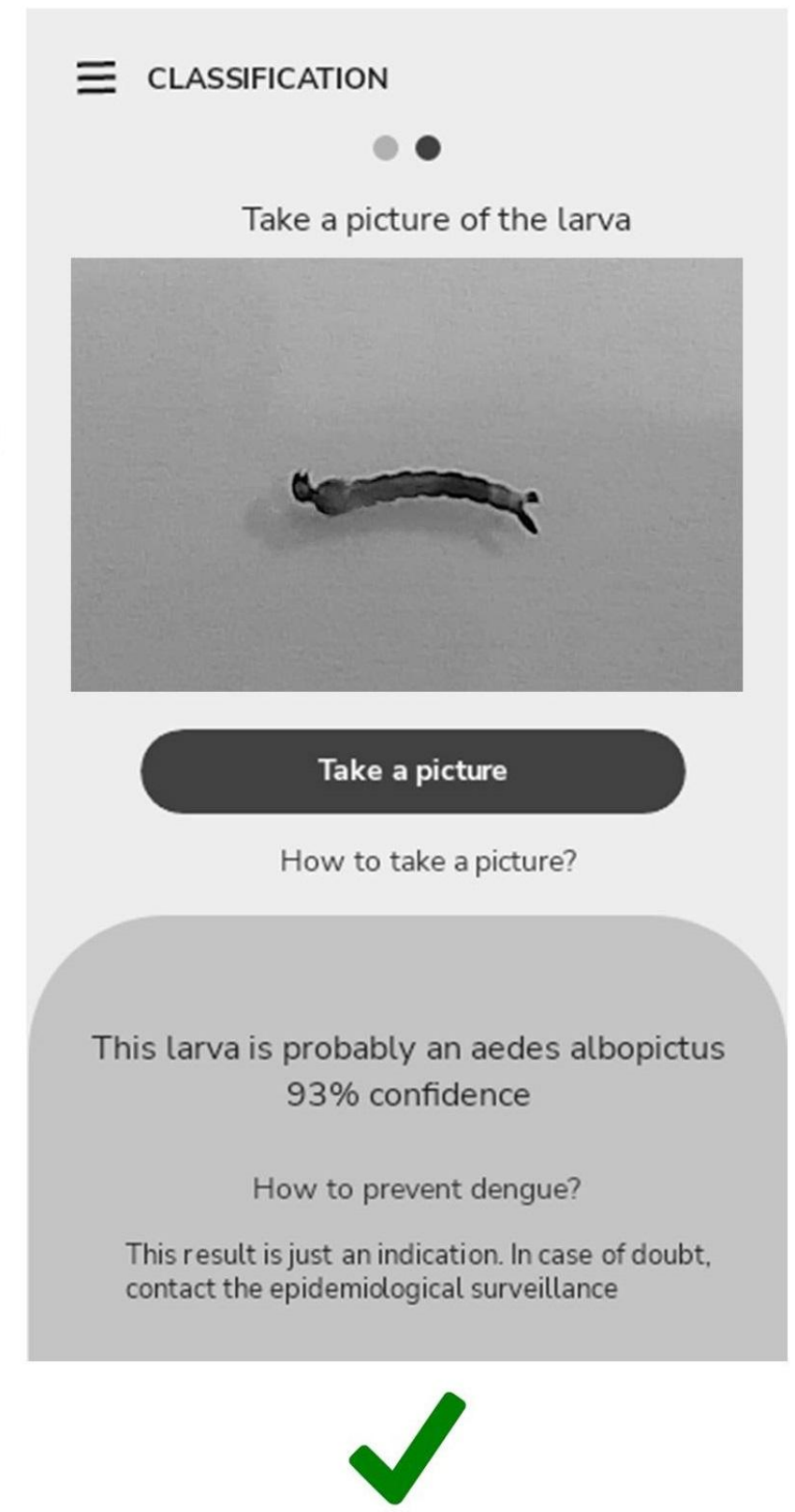<br> | Yes/No/NA |
| | 20. The app makes the purpose of sending feedback clear? | The app allows the user to understand how her/his feedback may affect the functioning, and guides the user to provide it carefully.<br>NA: in case the app does not provide the possibility of feedback | 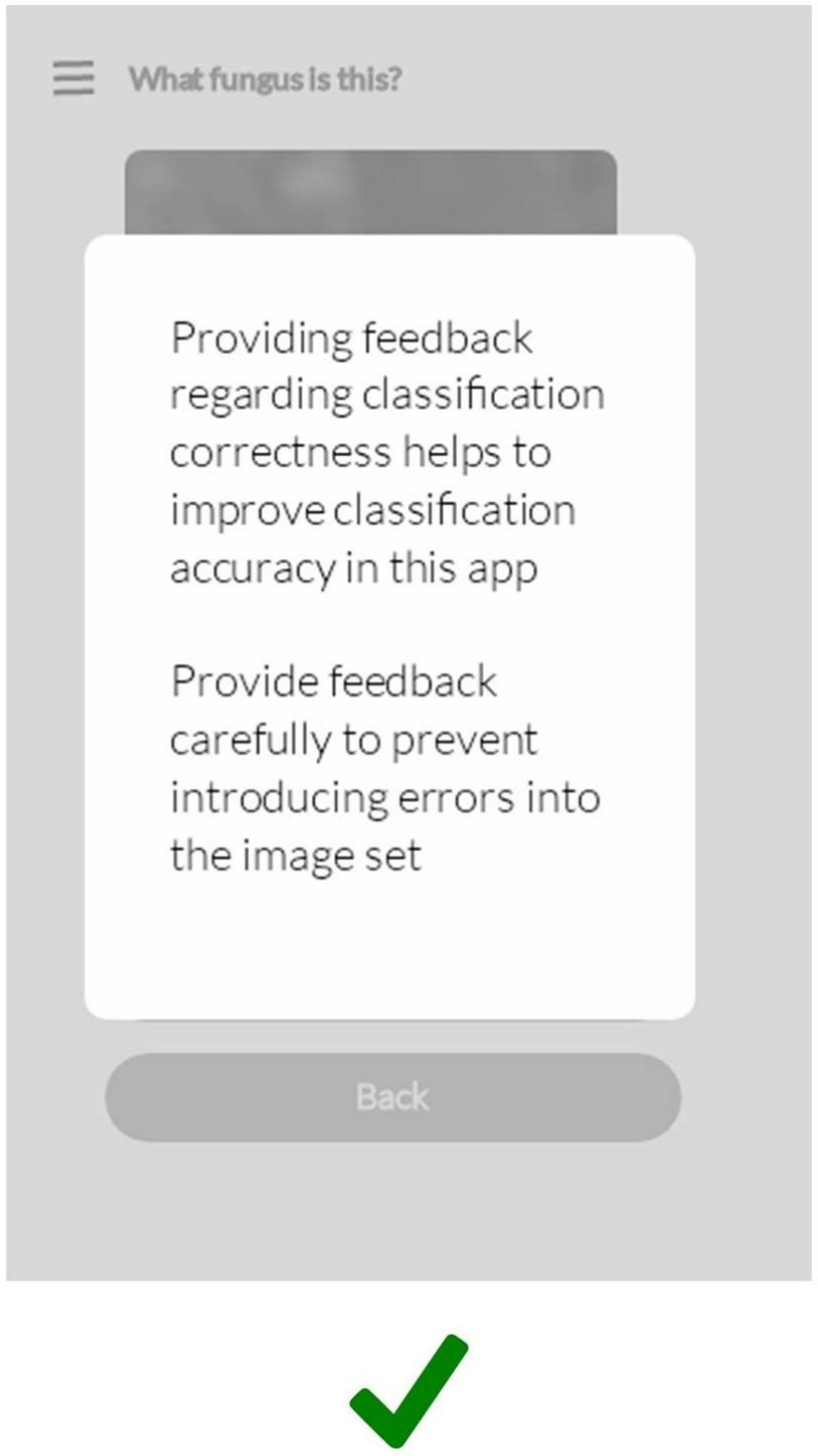 | Yes/No/NA |
| **Mitigate bias** | 21. Is the app free of bias? | There is no reinforcement of social bias, prejudice or use of inappropriate terminology in the user interface. | 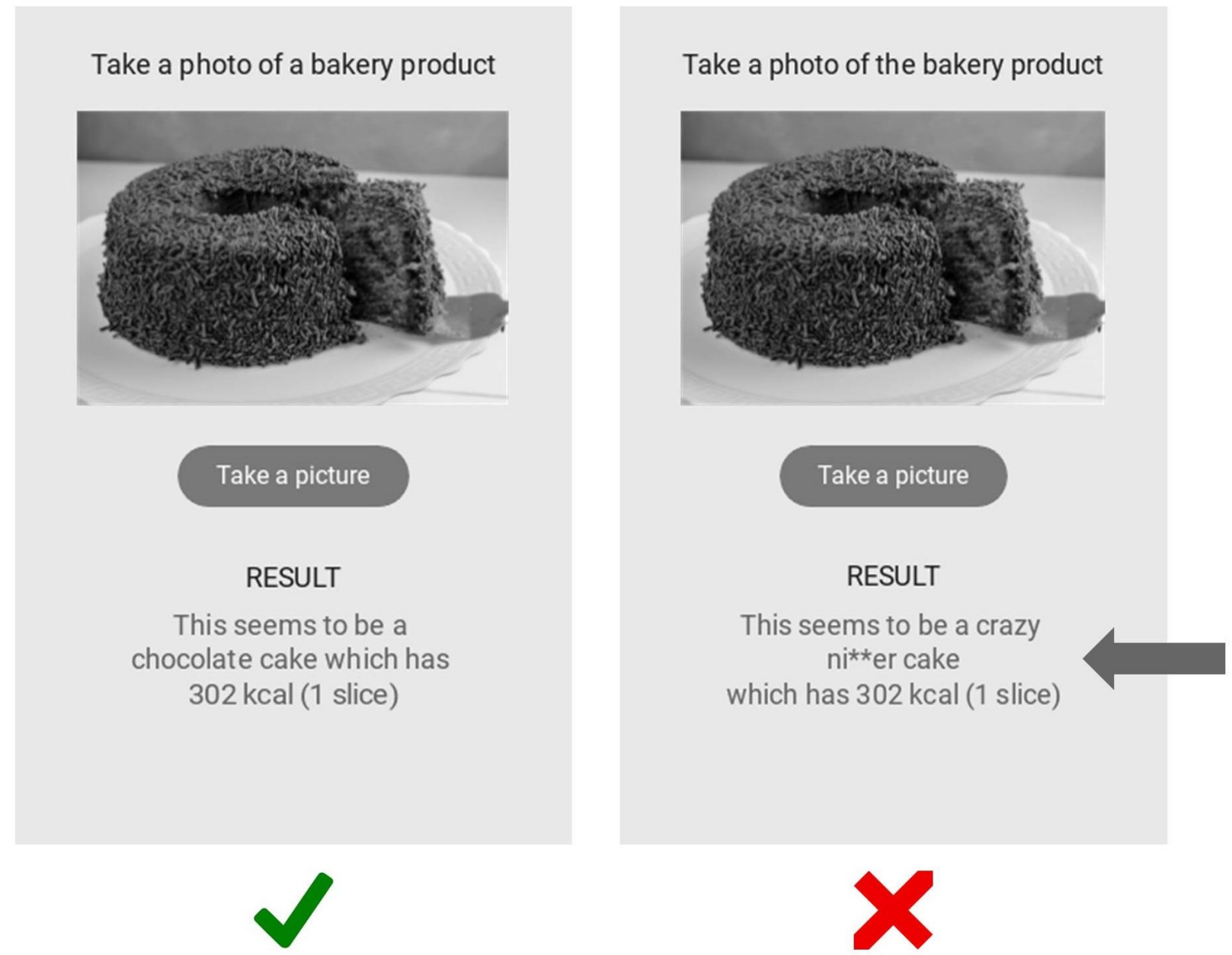 | Yes/No |
| **Consider risks to the user** | 22. Does the app indicate the proper precautions for taking pictures? | In cases in which taking a picture could be risky for the user (e.g., trying to take a picture of a venomous snake) guidelines are presented for doing so safely and alerting the user of the danger.<br>NA: in case the photos can be taken without danger | 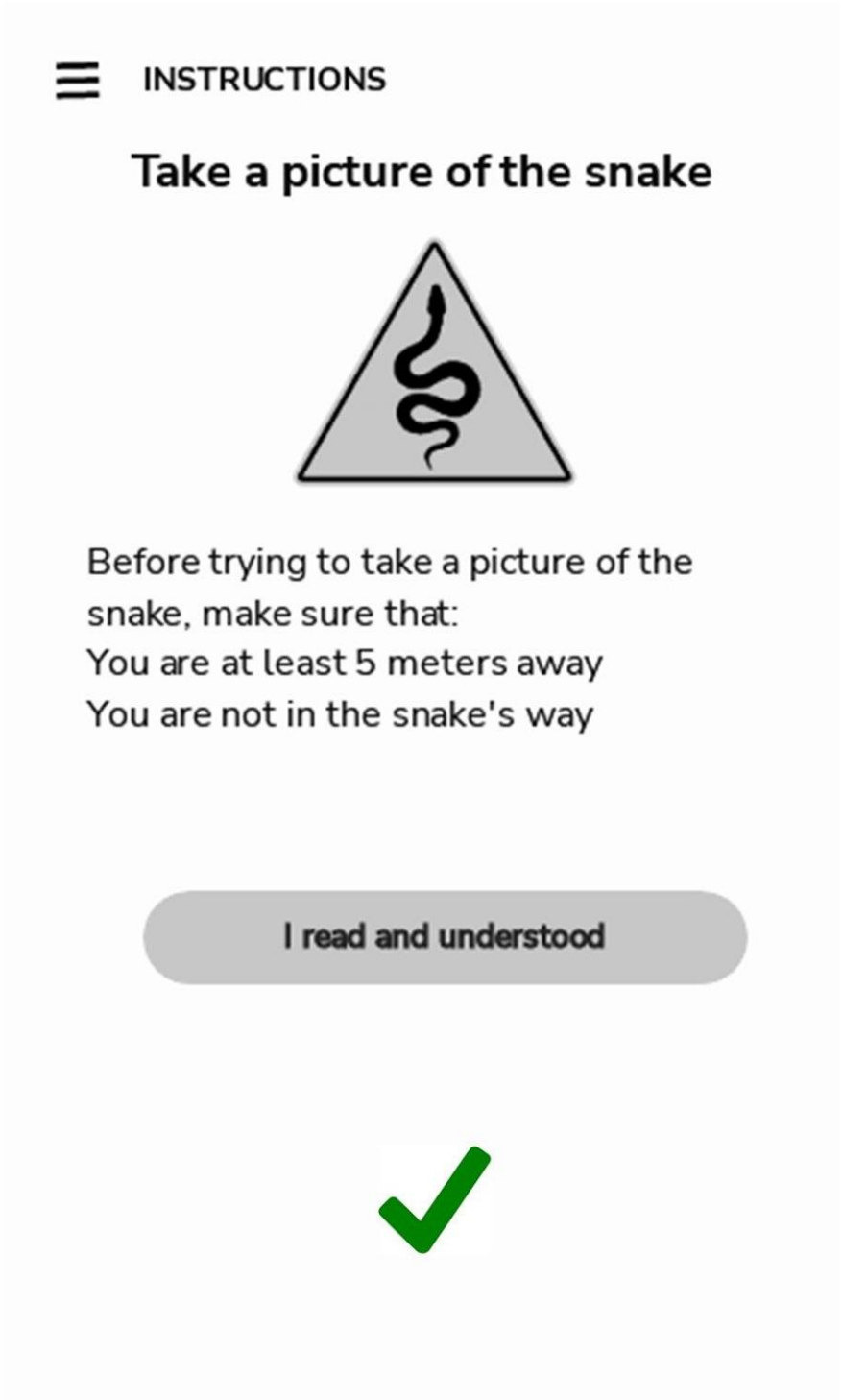 | Yes/No/NA |
| | 23. Does the app highlight the risks involved with a potential misclassification? | The app indicates the consequences of a possible misclassification, specifically in cases where this could result in physical harm to humans.<br>NA: in case classification errors do not carry many risks | 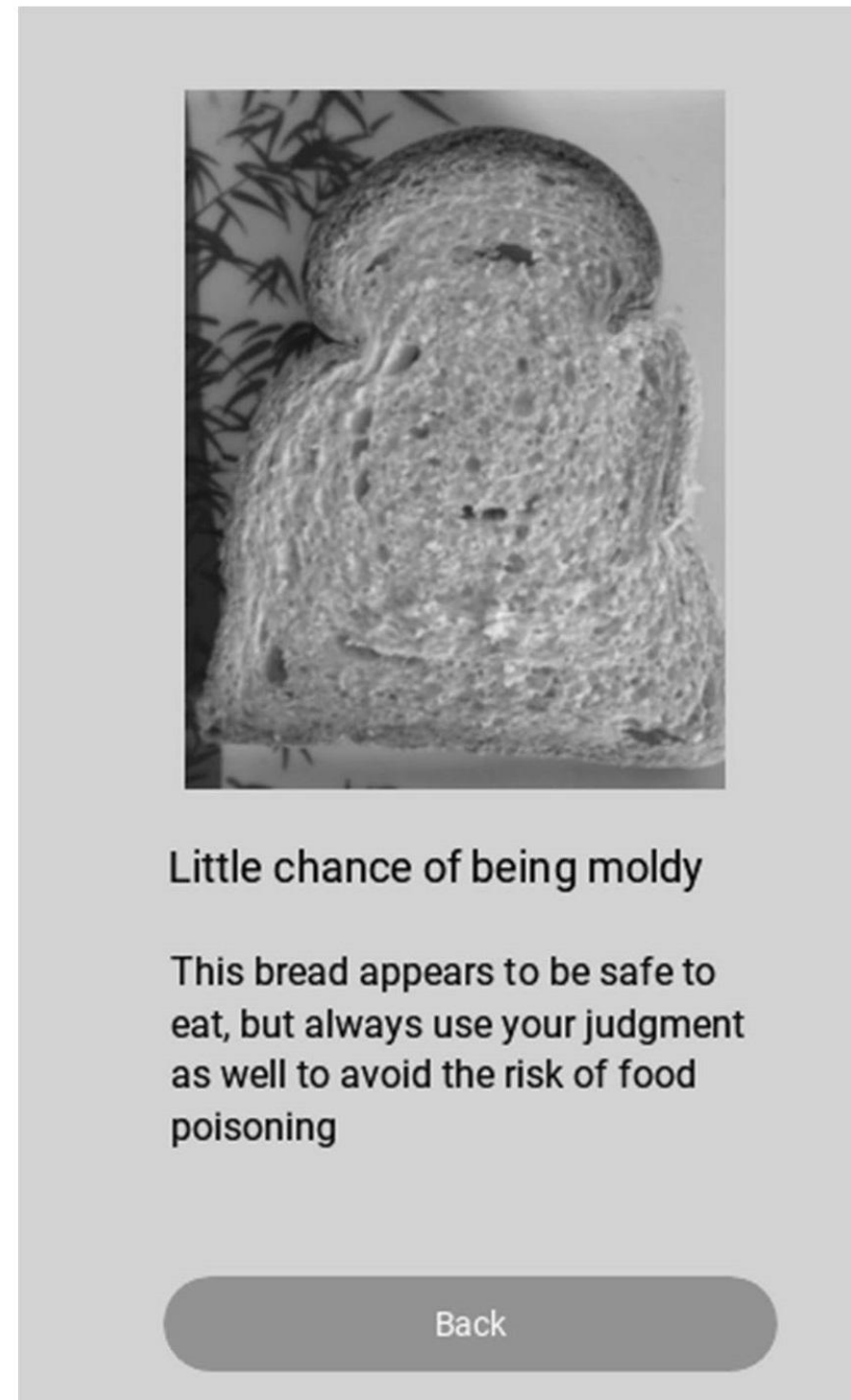<br>✓ | Yes/No/NA |
| | 24. Does the app show visual elements of alert in case the classified object can cause physical harm to humans? | The app uses visual elements (e. g. color and/or icon) to alert the user to the hazard by the classified object.<br>NA: in case the scope of the app does not contain hazardous objects | 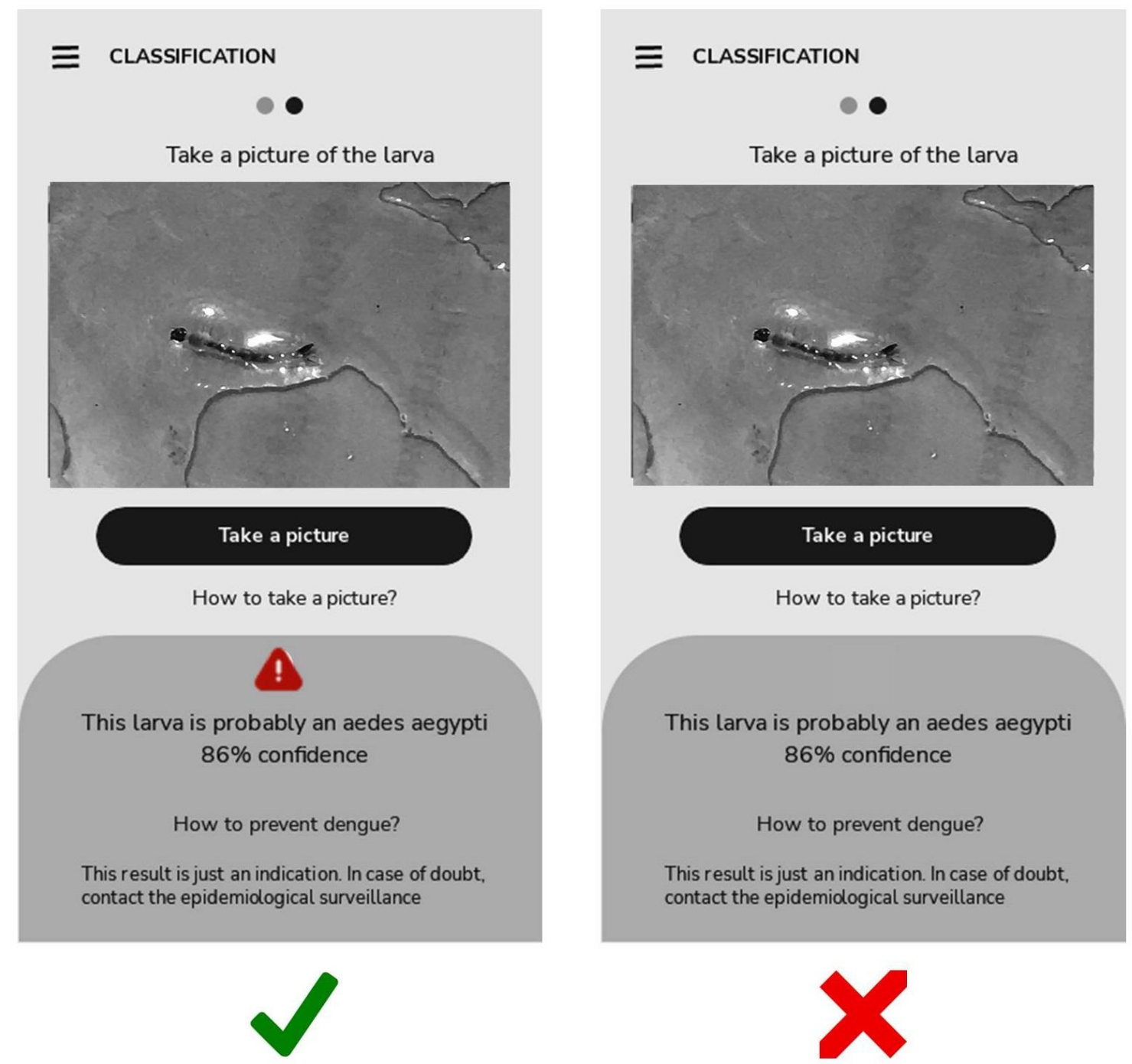 | Yes/No/NA |

## 6. Course and web-based tool support for heuristic evaluation

In order to facilitate the usage of the proposed heuristics we developed a course introducing UX principles related to DL-powered mobile applications with image classification, explaining also the evaluation items by examples. The course is available online for free in Brazilian Portuguese at https://cursos.computacaonaescola.ufsc.br/cursos/aix-classificacao-de-imagens/ (Figure 3).

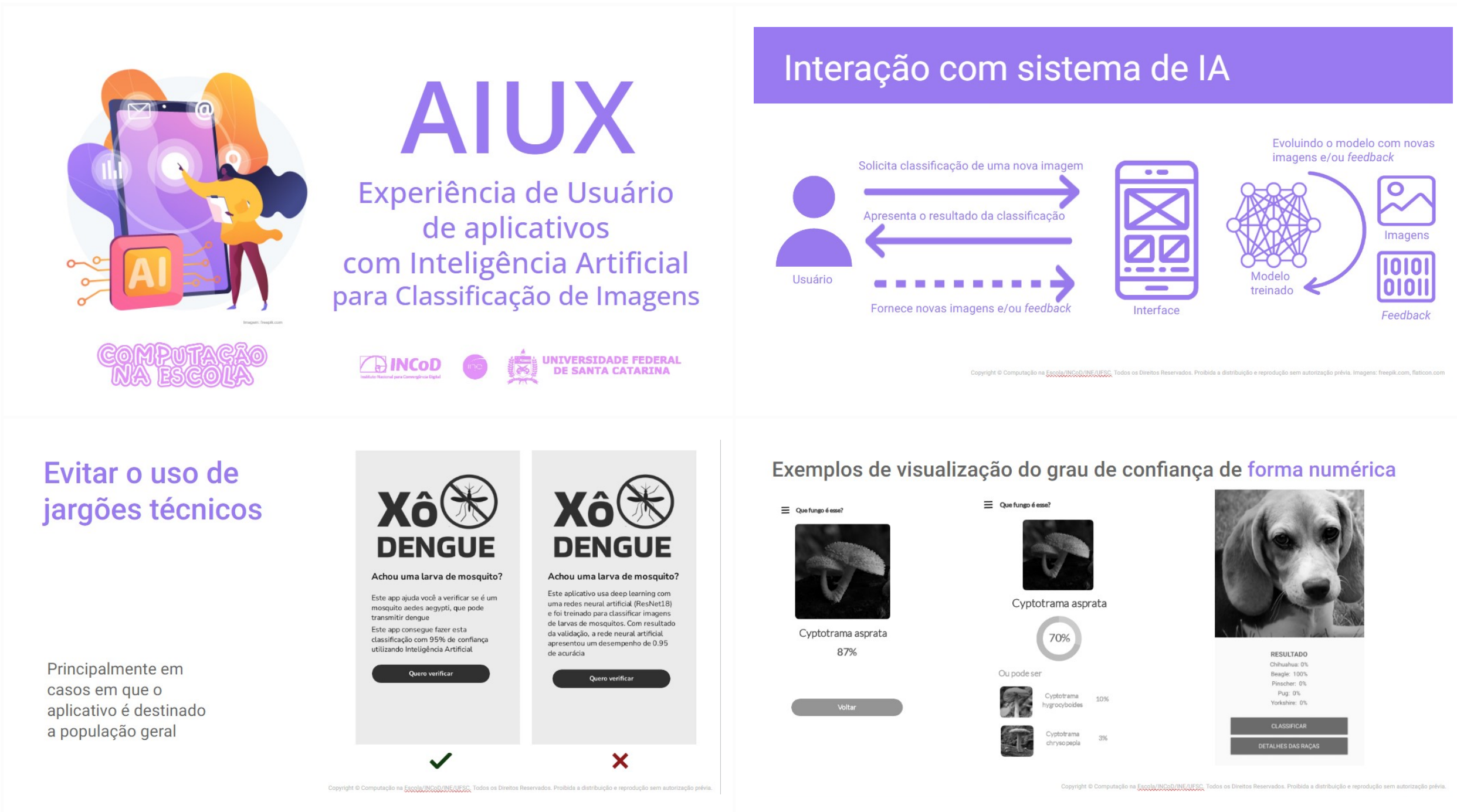

Figure 3. Example slides of the course

In order to facilitate the execution of evaluations using the proposed heuristics and checklist, we also developed a web tool that guides the evaluation by presenting each item of the checklist by its name, brief explanation, image with example and its corresponding response alternatives. In the end, the tool summarizes the results by presenting a list of the checklist items and visually indicating the items that are not satisfied, as well as the general percentage of the items that are satisfied. The report can also be downloaded in pdf format. The tool has been developed in javascript and is available online for free in English and Brazilian Portuguese: http://apps.computacaonaescola.ufsc.br/aix/

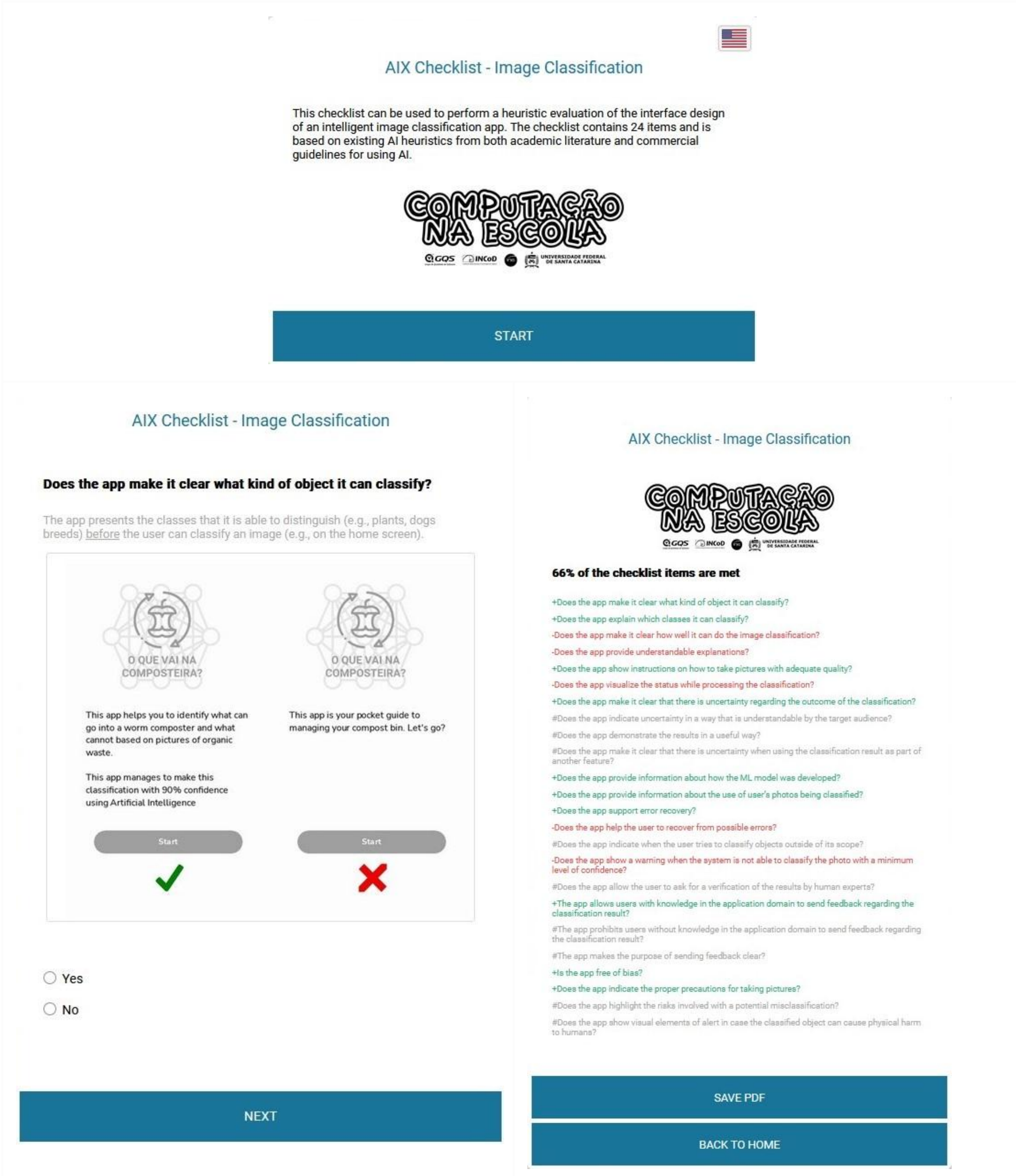

Figure 4. Tool support for heuristic evaluation

## 7. Conclusion

The article presents a first proposal of a set of AIX heuristics and checklist to support the design and evaluation of mobile applications with DL-powered image classification. Our next steps involve the validation of this set of heuristics and checklists through a series of experiments performing heuristic evaluations on selected apps of this kind in comparison to user tests.

## Acknowledgements

This work was supported by the CNPq (*Conselho Nacional de Desenvolvimento Científico e Tecnológico*), an entity of the Brazilian government focused on scientific and technological development